\documentclass[5p,times,usenatbib]{elsarticle}
\pdfoutput=1
\usepackage{ecrc}
\volume{00}
\firstpage{1}
\journalname{Astronomy and Computing}
\runauth{}
\jid{procs}
\jnltitlelogo{\includegraphics[width=1.5cm]{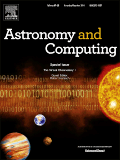}}
\usepackage{amssymb}
\usepackage[figuresright]{rotating}
\usepackage{tabularx}
\usepackage{booktabs}
\usepackage{graphicx}
\usepackage{blindtext}
\usepackage{multirow}
\usepackage{amsmath}
\usepackage[singlelinecheck=false]{caption}
\usepackage[ruled,vlined]{algorithm2e}
\usepackage{array}
\usepackage{tabu}
\usepackage{hyperref}
\graphicspath{ {images/} }

\makeatletter
\def\@author#1{\g@addto@macro\elsauthors{\normalsize%
    \def\baselinestretch{1}%
    \upshape\authorsep#1\unskip\textsuperscript{%
      \ifx\@fnmark\@empty\else\unskip\sep\@fnmark\let\sep=,\fi
      \ifx\@corref\@empty\else\unskip\sep\@corref\let\sep=,\fi
      }%
    \def\authorsep{\unskip,\space}%
    \global\let\@fnmark\@empty
    \global\let\@corref\@empty  
    \global\let\sep\@empty}%
    \@eadauthor={#1}
}
\makeatother

\begin{document}

\begin{frontmatter}

\dochead{}

\title{Evaluating virtual hosted desktops for graphics-intensive astronomy}

\author{Bernard F. Meade$^{1,2}$\corref{cor1}}
\ead{bmeade@unimelb.edu.au}
\cortext[cor1]{Corresponding author}

\author{Christopher J. Fluke$^{1,3}$}
\ead{cfluke@astro.edu.au}

\address{
$^{1}$Centre for Astrophysics \& Supercomputing,\\
Swinburne University of Technology, PO Box 218, \\
Hawthorn, Victoria, 3122, Australia}
\address{
$^{2}$Infrastructure Services (Doug McDonell Building), \\ The University of Melbourne, Victoria 3010, Australia}
\address{
$^{3}$Advanced Visualisation Laboratory, \\ Digital Research and Innovation Capability Platform, \\
	Swinburne University of Technology, PO Box 218,\\ 
	Hawthorn, 3122, Australia}

\begin{abstract}
Visualisation of data is critical to understanding astronomical phenomena. Today, many instruments produce datasets that are too big to be downloaded to a local computer, yet many of the visualisation tools used by astronomers are deployed only on desktop computers.  Cloud computing is increasingly used to provide a computation and simulation platform in astronomy, but it also offers great potential as a visualisation platform.  Virtual hosted desktops, with graphics processing unit (GPU) acceleration, allow interactive, graphics-intensive desktop applications to operate co-located with astronomy datasets stored in remote data centres.
By combining benchmarking and user experience testing, with a cohort of 20 astronomers, we investigate the viability of replacing physical desktop computers with virtual hosted desktops.  In our work, we compare two Apple MacBook computers (one old and one new, representing hardware and opposite ends of the useful lifetime) with two virtual hosted desktops: one commercial (Amazon Web Services)  and one in a private research cloud (the Australian Nectar Research Cloud).
For two-dimensional image-based tasks and graphics-intensive three-dimensional operations -- typical of astronomy visualisation workflows -- we found that benchmarks do not necessarily provide the best indication of performance.   When compared to typical laptop computers, virtual hosted desktops can provide a better user experience, even with lower performing graphics cards.  We also found that virtual hosted desktops are equally simple to use, provide greater flexibility in choice of configuration, and may actually be a more cost-effective option for typical usage profiles.
\end{abstract}


\end{frontmatter}



\section{Introduction}
Astronomy, as with many other scientific disciplines, is now in the petabyte-data era \citep{abello_massive_2002,kargupta_scientific_2008,juric_lsst_2012}.   This growth in the total volume of data is due, in part, to the improvements in resolution that modern instruments and detectors are able to access and record.  Alongside this is the increased computational power available for numerical simulations.

Visualisation is a crucial component of knowledge discovery. As both the size and complexity of astronomical datasets continue to grow, the existing paradigm of the astronomer visualising and analysing data at the desktop is being pushed to the limit.  The high computational and graphics-intensive requirements for many research workflows now exceed the processing, storage, and memory capabilities available with
standard desktop-based solutions \citep{berriman_how_2011,hassan_scientific_2011}.



A compelling option is to move all of the processing requirements away from the desktop to a dedicated remote data centre or into the cloud. Here, on-demand computational resources can be co-located with the data such that computation and analysis can be performed at an appropriate scale.

Choosing the right mix of dedicated compute resources that suit the needs of all users is complex.  The availability of cloud services  allows for flexibility and experimentation with configurations that is not always possible with a fixed-purpose data centre.  Cloud computing abstracts the hardware aspects of computing away from the user.  This takes away the burden of managing hardware, and allows the user to consume the service like a utility such as electricity or network connectivity. However, there is much that is still unknown, and untested, regarding the suitability, choice of hardware, cost effectiveness, and user experiences afforded by commercial and research clouds for supporting astronomical workflows.


To this end, the Square Kilometer Array (SKA) Telescope organisation\footnote{\url{https://skatelescope.org/}} and Amazon Web Services (AWS)\footnote{\url{https://aws.amazon.com/}} jointly announced the formation of the ``Astrocompute in the Cloud'' \citep[][]{astrocompute_seeing_2015} program in April 2015.  This program was proposed as a way to explore potential roles for AWS infrastructure to be used for current astronomy projects, and in the future for SKA-related research and operations.  This included opportunities to improve research outcomes through the application of additional on-demand compute power, storage and other  capabilities.

\subsection{The need for a virtual desktop}

Allocation and scheduling of computing resources through a prioritised batch queue is the preferred approach for most high-performance tasks.  For workflows that are computationally limited, any reduction in the overall processing time is beneficial.  The overhead in waiting for a workflow to be executed is amortised by the reduction in wall-time once the job starts.  However, many astronomy applications -- especially data visualisation tasks -- require an interactive, on-demand desktop window interface to operate.   Such an option is not always compatible with queued access to remote compute resources.

The paradigm of ``moving the computation to the data'' applies to both traditional computational tasks for analysis and knowledge discovery, and in the use of Virtual Hosted Desktops (VHDs; see \citet{miller_virtualization:_2007}). Here, the astronomer's virtual workspace resides entirely in the cloud, and is unlocked from the reliance on the processing capabilities of a physical desktop. A low powered local computer is only required as a gateway between user inputs (e.g. keyboard and mouse interaction) and streaming of images back to the display. Input requires minimal bandwidth; response speed is limited by the remote processing time and the overhead in returning image data to the display device, which scales linearly with the number of pixels.

VHDs can be provisioned on standard workstation or server computers, with the desktop environment presented as a Graphical User Interface (GUI) application management suite.  Alternatively the service can be installed on a virtual machine hosted by a cloud provider.

VHD capability has existed for some time, e.g. via X11 window forwarding, where graphics are remotely rendered and streamed via a connection protocol such as SSH, or through Virtual Network Computing \citep[VNC;][]{duato_interconnection_1997}.  These approaches are usually reserved for circumstances where the performance of the environment itself is not critical.  In this way, VHDs have often been used as a last resort due to their bandwidth and latency issues.  With the rapid improvements in modern networking, it is now possible to employ a VHD in a manner that is almost indistinguishable to the local desktop.

For graphics-intensive work, the result of using a VHD has not always been satisfactory. This, too, has changed with the advent of graphics
processing unit (GPU) acceleration of remote desktops. Indeed, GPU manufacturers such as NVIDIA are creating graphics cards specifically for operation in cloud infrastructures (e.g. NVIDIA Grid K1 and M10)\footnote{\url{https://www.nvidia.com/en-us/gpu-cloud/}}.  It is timely, therefore, to explore whether a virtual hosted desktop is a functional
replacement for a local computer in astronomy.




\subsection{Overview}
\label{lbl:overview}

In this work, we investigate the suitability of VHDs for performing  visualisation tasks from the domain of astronomy. Combining benchmarks with user experience testing \citep{lam_empirical_2012} through the involvement of a cohort of astronomers, we compare software performance and user experiences between local computers (two generations of Apple Mac laptops) and two VHDs, provisioned by AWS and the Australian National eResearch Collaboration Tools and Resources (Nectar)\footnote{\url{https://nectar.org.au/research-cloud/}} Research Cloud (NRC). These options cover three potential choices for upgrading a computing environment for use in graphics-intense workflows: buy a new local computer; purchase time through a commercial cloud; or, if the option is available, utilise a national research cloud infrastructure.

One of our motivations is to provide 
astronomers with the knowledge to make more informed decisions when it comes to investing in either a new physical desktop or a VHD.  Choosing an alternative infrastructure requires a consideration of operational factors, suitability, user experience, and financial matters.  While subject to change without notice, we compare pricing models (at the time of writing) for both physical hardware and cloud services.


This remainder of this paper is laid out as follows.  A review of the previous work done in this research area and an introduction to VHDs is presented in Section \ref{lbl:background}. In Section \ref{lbl:experiment}, we benchmark two laptops and two VHDs, and describe the user experiences in these environments for 2D and 3D astronomy tasks. In Section \ref{lbl:discussion} we discuss the results of the user experience testing, and compare the costs associated with VHDs and laptops.  Concluding remarks are made in Section \ref{lbl:conclusion}.

\section{Cloud computing and virtual hosted desktops}
\label{lbl:background}

\subsection{Cloud computing in astronomy}
Cloud computing allows clustered commodity computers to be provisioned in the form of Product-as-a-Service, where the product being consumed could be a database (DBaaS), a development platform (PaaS), or most commonly, infrastructure (IaaS). Most commercial cloud services provide a mix of these options.  On allocation of the resource, the virtual instances can be used to perform a variety of tasks, ranging from scientific computation to  running web servers. 

Cloud computing for scientific workflows has been investigated by several groups over the last decade or so \citep[e.g.][]{deelman_cost_2008, hoffa_use_2008, juve_scientific_2009}.  One of the first investigations into the use of cloud specifically in astronomy workflows was published by \citet{berriman_application_2010} and was extended with more detailed benchmarking in \citet{vockler_experiences_2011}.  These papers showed that commercial clouds, in these cases AWS, could be used cost effectively to provide substantial ad-hoc computation resources.  

\citet{ball_canfar+_2013} described data mining with the machine learning platform, SkyTree\footnote{\url{http://www.skytree.net/}}, running on CANFAR\footnote{\url{http://www.canfar.net/en/}}, the cloud computing platform for the Canadian Astronomy Data Centre\footnote{\url{http://www.cadc-ccda.hia-iha.nrc-cnrc.gc.ca/en/}}.  This research established that cloud computing was a viable option for certain types of computation in astronomy, in this case, the data mining of a 13 billion object catalogue.

Beyond on-demand computation, cloud computing can provide a suitable platform for visual tools.  The Montage Image Mosaic Toolkit, as mentioned in \citet{deelman_cost_2008} and \citet{hoffa_use_2008}, was used on AWS to create a Galactic Plane atlas, which combined data from the 2MASS, GLIMPSE, MIPSGAL, MSX and WISE sky surveys \citep{berriman_application_2013,berriman_next_2016, berriman_application_2017}.  These studies found that cloud infrastructure provided increased flexibility in resource provisioning, reducing initial financial outlay and costs overall.  However, an increased understanding of service models and cloud resource management was required to effectively use cloud services.  For applications such as Montage, with short job runtimes, the cloud approach provided good compute resource utilisation, while longer, more computationally intensive jobs were less cost-effective.  There is also the risk of resource availability and network connectivity introducing unexpected and indeterminate delays.
 
Cost-benefit analyses have been conducted in relation to the use of cloud with the SKA pathfinders, such as LOFAR\footnote{\url{http://www.lofar.org/}} and CHILES\footnote{\url{http://chiles.astro.columbia.edu/}}. \citet{sabater_calibration_2017} used cloud infrastructure to run the LOFAR calibration pipeline, finding the flexibility and ad hoc availability of the cloud provided a better option than traditional on-premise HPC services. \citet{dodson_imaging_2016} conducted direct comparisons of the CHILES imaging pipeline using a local cluster, a National Peak cluster (Magnus at the Pawsey Supercomputing Centre\footnote{\url{https://www.pawsey.org.au/}}, Western Australia), and cloud infrastructure from AWS.  For both the LOFAR and CHILES projects, the cloud platforms were found to be highly competitive across most measures, where costs such as operations are offset against capital expenditure on a local cluster.  

A more general discussion of the taxonomy of cloud service providers can be found in \citet{antonopoulos_taxonomy_2010}.  Further discussion of cloud, high performance computing and big data, including the impact of virtualisation can be found in the PhD Thesis of \citet{younge_architectural_2016}.


\subsection{Virtual hosted desktops in astronomy}
The provision of a VHD is achieved by connecting a local computer to a remote server, which appears on the local computer as a desktop, with all the pre-installed applications ready to use.  A thin client \citep{nieh_comparison_2000, deboosere_cloud-based_2012} is so called because it requires the local client computer to perform very little computationally, while the power to drive the application comes from the server the client is connected to.  A common method of connection is via VNC 
which uses the Remote Desktop Protocol \citep[RDP;][]{khalid_desktop_2016}. Many astronomers are familiar with telescope operations being managed using VNC \citep[e.g.][]{caton_remote_2009}.  More general detail about Desktops-as-a-Service can be found in \citet{bipinchandra_intelligent_2014} and \citet{khalid_desktop_2016}.

An early example of the use of cloud-based desktops in astronomy is detailed in \citet{berriman_tale_2012}.  During the 2012 Carl Sagan Workshops\footnote{\url{http://nexsci.caltech.edu/workshop/2012/}} hosted by NExScI\footnote{\url{http://nexsci.caltech.edu/}}, Berriman and his team successfully used AWS' Elastic Compute Cluster (EC2) service to provision VHDs for use in training 160 astronomers to reduce and analyse Kepler light curves.  Rather than have the participants install and configure the raft of applications required for the workshop, a pre-installed suite was available to connect to via VNC.

Some national peak facilities provision Desktops-as-a-Service as a means to access computation and storage services. ACID (Astronomical and physics Cloud Interactive Desktop) is a suite of desktop applications for astronomers and physicists provided by the Cherenkov Telescope Array (CTA)\footnote{\url{https://www.cta-observatory.org/}} Science Gateway (Italy) for the research community \citep{massimino_acid_2014}.  This service provides a VNC User Interface which is accessible through a web browser.  This approach eliminates the need for a VNC client installation on the local device.  


Many astronomy applications require three-dimensional (3D) graphics acceleration.  Services like MASSIVE\footnote{Multi-modal Australian ScienceS Imaging and Visualization Environment \url{https://www.massive.org.au/}} augment cloud services with GPUs to support these applications \citep{goscinski_massive:_2015},  usually via a ``pass through'' model, where each GPU is used to support a single virtual machine \citep{ravi_supporting_2011}.   This allows a virtual machine to direct OpenGL calls to the physical GPU on the host, allowing the virtual machine to run GPU-dependent applications, with very little impact from virtualisation.

It is now possible, though not necessarily widely available, for cloud services to virtualise the GPUs in the hosts.  For example, a GPU existing in one node can be shared with several of the virtual machines running on that host.  Like the core utilisation, the virtual GPU is managed to respond to GPU requests coming from each virtual machine \citep{cardoso_closer_2016}.  This approach can be extended to provide network access to virtual GPUs, allowing virtual machines on other hosts to access GPU resources.  This is often used to support general purpose GPU computation for numerical calculations, but can also support GPU integration to VHDs \citep{hong_gpu_2017}.

\subsection{Comparing cloud and physical desktops}
\label{lbl:clouddesktops}

A physical computer, in the form of a desktop workstation or laptop computer, is an essential tool for modern astronomy.  It is important to ensure that any replacement, such as a VHD, is capable of providing a better value proposition. 
A dependence on computers means that many astronomers have a high-level of technical computing competence, and are often quite particular when it comes to choosing IT hardware.  When purchasing a computer, factors such as number of computer cores, clock speed, RAM and GPU capabilities are all important in making a decision.

The same is true for choosing a virtual machine, whether it is to be used as a VHD or not.  However, choosing a virtual machine from a cloud provider typically offers far more potential for customisation, unless the researcher is willing and able to build a physical computer from parts.  More importantly, making a mistake is far easier to correct with a virtual machine, as the chosen options can be discarded, and a new specification built in its place.

If a virtual hosted desktop is to be considered a viable alternative to a local desktop for graphics-intensive workflows, it must meet certain criteria:
\begin{enumerate}
    \item It must be a simple process to use the environment.  If astronomers find it difficult or impractical to use, or requires significant education to learn how to use, then it is unlikely to be adopted by the community.
    \item It must be a smooth experience.  Even if the process to use the environment is simple, it must be able to run astronomical and related software smoothly. Low-responsiveness to keyboard and mouse movements, and delays in running applications and loading data will result in a frustrating experience for astronomers.
    \item It must be demonstrably cost-effective. Many researchers do not wish to change from an environment that they are familiar with, but if a solution can be shown to be cost effective, or show significant benefits in other tangible ways, they may be more willing to explore it.
    \item It must be powerful enough to do the tasks required.  Modern astronomy workloads are increasingly demanding, either computationally, data intensively, or visually.  The ability to choose a fit-for-purpose compute capability for a specific task is one of the main attractions of cloud computing.  If the selected environment is under-powered, a more powerful option can be selected next time.  If the environment is more powerful than required, a lower-powered option can be selected.  Flexibility means the right resource is available when required, and can be relinquished when it is not.
    \item It must be available when required.  Astronomers, like most researchers, are turning to portable devices for research and other work.  The portability of modern devices allows for work to be conducted in non-traditional settings.  This has led to an expectation that research can be conducted anywhere and at anytime.  However, a cloud based service has an increased dependency on network connectivity, which may not always be present or sufficient.
\end{enumerate}

In the next section, we describe the performance of VHDs using both quantitative and qualitative measures, by comparing system specifications and benchmarks result, as well as participants' reactions to the environments.

\section{User experiences with virtual hosted desktops}
\label{lbl:experiment}

\begin{table*}
	\centering
	\caption{Technical specifications for the four computing environments used for benchmarks and user experience testing.}
	\small
	\begin{tabular}{ p{11.75em} p{7em} p{14.75em} p{1.75em} p{5cm}} 
		\hline
		Environment & Operating system & CPU/Cores & RAM & GPU\\
		\hline
		MB13 & MacOS 10.10.5 & Intel i7-3740QM CPU(2.70GHz) x 4 & 16GB & NVIDIA GeForce GT 650M (1024MB)\\
		MB17 & MacOS 10.10.11 & Intel i7-6700HQ CPU(2.60GHz) x 4 & 16GB & AMD Radeon Pro 450 (2048MB)\\
		NRC VM (mel.gpu-k1.large) & Ubuntu 14.04 & Intel Haswell CPU(2.30GHz) x 4 & 16GB & NVIDIA K1 GPU (256MB)\\
		AWS VM (g2.2xlarge) & Ubuntu 14.04 & Intel Xeon CPU(2.60GHz) x 8 & 15GB & NVIDIA GRID GK104 GPU (256MB)\\
		\hline
	\end{tabular}
	\label{tab:systemspecs}
\end{table*}

To date, the majority of the discussion on the usefulness of virtual hosted desktops is based on anecdotal evidence and supposition.  To remedy this, we recruited 20 astronomers to participate in user experience testing.  This is an established approach to testing the suitability of an environment, application or interface that depends on the effective use by human operators. User experience testing methods include informal evaluation, observations of how software or systems are used -- in the field or in controlled settings -- and questionnaires [see the taxonomy of evaluation methods in \citet{lam_empirical_2012} in the context of information visualisation].  While mindful of limitations based on the number, and experience levels, of participants, user experience methodologies recognise there is value in obtaining immediate, subjective responses from even a small cohort of users.  User experience testing approaches can be used in conjunction with more objective measures, such as bench-marking data, to improve the value outcomes when purchasing computational resources \citep{bevan_usability_2009,rampersad_improving_2017}.

During the user experience tests, we record two main types of data:
\begin{enumerate}
\item Quantitative. By timing how long participants take to complete a task or monitoring frame rates during a task, we gain insight as to which classes of tasks are suited to each of the desktop environments. 
\item Qualitative. By asking astronomers to perform typical visualisation tasks with current astronomy software, we are able to use the reflections of the participants to gauge the human experience factors of VHDs.
\end{enumerate}

\subsection{Properties of local and virtual desktops}
\label{lbl:systems_specs}
Consider two common purchasing scenarios faced by astronomers:
\begin{enumerate}
    \item I am a new staff member or postgraduate research student. What is the best standard option I can access?  In this case, a new device is anticipated.  
    \item My computer is old, can I get a replacement?  In this case, the device in question may be out of warranty and considered suitable for replacement under standard university IT renewal schemes.
\end{enumerate}
For our user experience testing, we selected four different computing options that addressed such purchasing scenarios.

As many research institutions use a 3 to 4 year renewal cycle for their computer fleet, a MacBook Pro 2013 (MB13) and a MacBook Pro 2017 (MB17) were used to represent the two life-cycle edge cases of a physical desktop environment. They were chosen because they represent the mid-to-high-end options for researchers purchasing a new laptop in 2013 and 2017.   Both MacBooks were equipped with graphics accelerators and have a screen resolution of 2880 $\times$ 1800 pixels.  

For comparison with these physical desktops, we chose to investigate VHDs offered by one commercial cloud -- Amazon Web Services because it is available worldwide, has competitive pricing structures, and a large selection of infrastructure options -- and one national research cloud service -- the Nectar Research Cloud, which is a private option, available to all Australian researchers.

\subsubsection{Amazon Web Services}
\label{lbl:aws}
Public cloud providers like AWS offer a fully online service where anyone with a credit card can sign up and start a virtual machine in a matter of minutes.  

AWS provides a wide range of pay-for-use computation and data resources accessible over the Internet, along with a number of managed services.  The IaaS mode is the most common use, where users request a virtual machine with specific characteristics, such as number of compute cores, amount of RAM, amount of attached storage, and an operating system.  A virtual machine is launched on AWS infrastructure in one of several data centres located around the world, and made available to the user via the Internet.  

The user connects to the virtual machine via certain protocols, most commonly SSH, and can install any required software.  If a windowing (i.e. desktop) environment is installed, the user can also configure the virtual machine to allow VNC connections.  With a local desktop VNC client, the user can then connect to the virtual machine desktop and operate the virtual machine as if it were their local desktop, with local keyboard and mouse activity being passed through to the virtual machine.

A full discussion of the pricing models of AWS is beyond the scope of this paper, however, two options are relevant: 
\begin{enumerate}
\item {\em On-demand pricing} is the simplest to use and plan for, but is also the most expensive.  Using this model, a fixed price for a virtual machine configuration is known and agreed to before the virtual machine is created.  
Importantly, the On-demand price guarantees the availability of the virtual machine while it is being used.  This provides surety and clarity when planning the actual usage of the resource.

\item {\em Spot pricing} allows AWS to sell the unused capacity in its data centres at a far more attractive rate than the On-demand price.  However, as demand on that resource increases, the Spot price will rise.  Once the Spot price exceeds the user's bid price, the user's virtual machine will be stopped.  The user will need to increase their bid to allow the virtual machine to be restarted.  
\end{enumerate}

In this work, we only use AWS On-demand instances, as interactive visualisation workflows require guaranteed and continuous access to the VHD. 

\subsubsection{Nectar Research Cloud}
\label{lbl:nrc}
Many research institutions or federations offer private research clouds specifically for their research communities.  In Australia and New Zealand, Nectar established the largest private Research Cloud in the Southern Hemisphere.  Private research clouds are not as big as public cloud providers like AWS, but they are generally more suited to the demands of research.

Operating in a similar way to AWS, the Nectar Research Cloud is an Australian Federal Government initiative, which commenced operation in 2012. It is designed to support the computation and storage needs of the Australian research community using a federated private research cloud \citep{meade_research_2013}.
The NRC offers over 32,000 cores, distributed between nine physical nodes, and has supported in excess of 10,000 users from the Australian research community.  

Access to the NRC is either through a research merit application or via a host institution's private infrastructure.  Researchers are not directly charged for their use of the NRC under the merit scheme, and institutions determine their own access model for their private resources.  NRC primarily provides IaaS to the Australian research community, though new services continue to come online as the service matures. 

The Melbourne Node of the NRC, hosted at the University of Melbourne, offers limited GPU capability for VHDs.  A mix of NVIDIA K1s and M10s is provided (though the M10s were not available during our user study), with the GPU-enabled hosts operating in a ``pass-through'' configuration to the virtual machine.  

\subsubsection{Technical specifications}
Table \ref{tab:systemspecs} summarises the system specifications for the MB13, the MB17, the NRC virtual machine and the AWS virtual machine. 

The cloud virtual machines were chosen based on availability.  At the time of the investigation, the AWS virtual machine g2.2xlarge was the cheapest of the available options.  The more expensive g2.8xlarge provided considerably more CPU computation power than either of the local computers, and so was not used.  The NRC virtual machine chosen was the most closely matched to the AWS virtual machine instance from the available flavours.  During the study, the virtual machine environments were run in full screen mode to match the graphical rendering load of the local laptop screen.

Due to licensing requirements from Apple Computer PTY LTD, the Macintosh operating system, MacOS X, is not able to be used in a cloud environment. While the Microsoft Windows license does allow for use in cloud environments, it does not support some of the applications needed for the investigation, so the selected operating system for the VHDs was Ubuntu Linux 14.04 LTS (Trusty Tahr).  The MB13 was used at Swinburne University of Technology and the MB17 was used at the University of Melbourne. TurboVNC\footnote{\url{https://www.turbovnc.org/}} was used to connect to the VHD.  When operated in full screen mode, TurboVNC automatically adjusts screen resolution according to the size of the attached display, which is 2880 $\times$ 1800 for both the MB13 and MB17.

A network with sufficient bandwidth and stability is critical to providing a persistent connection to the virtual machine supporting the VHD. Both the University of Melbourne and Swinburne University of Technology have substantial network infrastructure, both wired and wireless.  At Swinburne University of Technology, the network used was {\em Eduroam}\footnote{\url{https://www.eduroam.edu.au/}}, a federated wireless research network with peering institutions all around Australia.  Six of the University of Melbourne participants used the University's wired 1Gbps network, while the remainder used the University's wireless solution, {\em uniwireless}. The network performance was measured before and after the tasks using the Speedtest website\footnote{\url{http://www.speedtest.net}}, to ensure the network was stable throughout.  

To minimise the impact of network latency during the user experience testing, the AWS data centre located in Sydney Australia, was chosen as this is the closest option to the University of Melbourne and Swinburne University of Technology. 

Results from these networks tests are shown in Appendix \ref{lbl:apndx-network}.

\subsection{Benchmarking the environments}
\label{lbl:benchmarking}

Computing products are released with technical specifications, which are usually considered objective measures of a product's performance in certain conditions.  However, computational environments are complex, and individual components might not be operating in ideal conditions, resulting in less than optimal performance.  To accurately determine the true performance of a complete system requires the performance measurement to be conducted under the conditions of intended use.

\begin{figure}
	\includegraphics[width=\columnwidth]{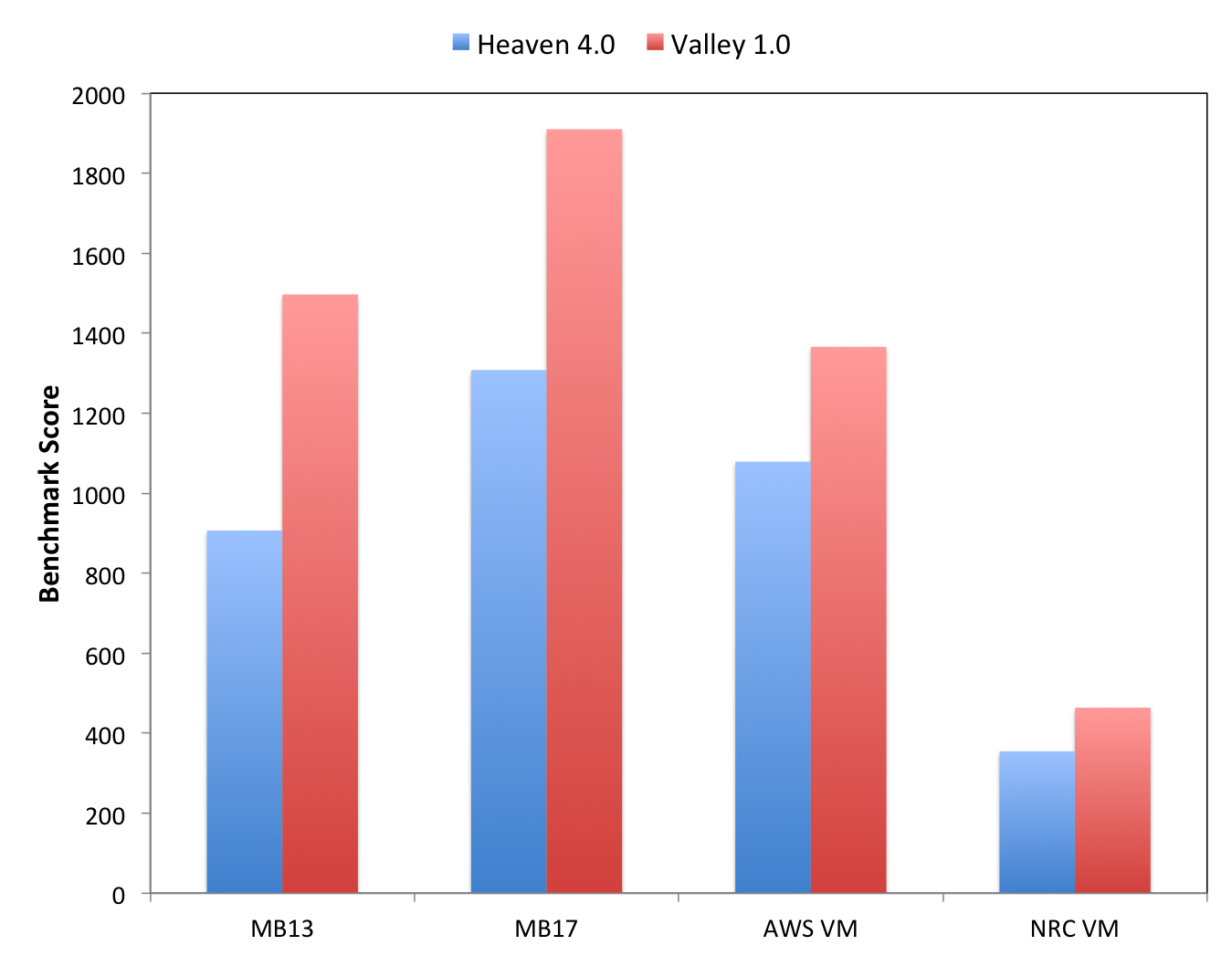}
    \caption{Scores for the Unigine Heaven 4.0 and Valley 1.0 Basic benchmarks obtained for each of the four computing environments. The Valley benchmark includes improvements on the older Heaven benchmark. The score is calculated from a combination of the maximum, minimum and median frame rates, as well as CPU and GPU performance.  This graph indicates the MB17 is the most performant system, while the NRC VHD is the least performant.}
    \label{fig:UnigineBenchmarks}
\end{figure}

The most common way to compare systems' performance is to run benchmarks.  Benchmarks are only useful if they test a system's capacity in conditions that fully expose the system's limits.  While many benchmark options are available, not all are suited to the situation being investigated here.  The Unigine Valley and Heaven benchmarks\footnote{\url{https://benchmark.unigine.com/}} were chosen for heir cross-platform availability, free accessibility and use of highly demanding 3D graphics computation.  The Heaven benchmark tests vertex and texture operations as well as lighting effects, while Valley, released later, expands these to include atmospheric effects and performance optimisations.  Figure \ref{fig:UnigineBenchmarks} shows the results for each of the four computing environments, with the scores being calculated based on CPU and GPU performance, and maximum, minimum and median frame rates.  These results are intended for comparison between the four systems in this investigation, rather than an independent objective measure against any system.

The local computers performed very well for the Unigine Valley and Heaven benchmarks. The AWS virtual machines performed slightly better than the MB13, but the NRC virtual machine was easily the lowest.  This partially aligns with the expected performance based on the currency of the GPU, with the MB13 being the oldest, the AWS GRID GK104 being next, and the MB17 being the newest.  The NRC K1 is approximately the same age as the MB13, but is not as performant.  In each case, the predicted performance does align with the benchmark results.

Benchmarks provide a valuable method of comparing systems out of context when in context comparison is not available.  For the most part, this is an accepted approach to determining the potential suitability of a system for a task that is similar to the benchmarking application.  However, this is less useful if the nature of the intended task is uncertain, or exposes an unanticipated demand on the system.

Furthermore, a value choice might be informed by benchmark results, but factors such as user experience must also be taken into consideration.  For example, there is little value in buying the best graphics card if users are unable to distinguish the difference when compared with a lower performing card.

Having completed benchmarking, we investigated whether user experiences with the physical and virtual hosted desktops were consistent with a benchmark-only approach.

\subsection{Participants}
\label{lbl:participants}
Because the focus in this investigation is the astronomers' experience, the participant cohort was limited to astronomers, either current academic staff or postgraduate research students (including recently graduated students).  Ten participants were recruited from the astronomy department at Swinburne University of Technology and a further ten from the University of Melbourne.  Three participants were academic staff and 17 were postgraduate students.  

No previous experience of the software or techniques involved was required to complete the tasks.  Where previous experience with the software was identified, these participants were encouraged to adhere to the instructions provided, even if they differed from their usual practices.  

Each participant was asked to complete a brief interview before the hands-on component of the user investigation was undertaken.  This survey was designed to understand the cohort's collective experience of cloud computing in general and VHDs in particular.

From this survey, 60\% of the cohort could provide a reasonable definition of cloud computing, with the remaining 40\% either unsure or confused it with online storage. 80\% were able to provide a reasonable definition of a VHD, though only 20\% had used VHD in their research, compared to 60\% who had used the cloud (including cloud storage) for their research.  

17 of the participants had used X11 Forwarding or VNC (typically for telescope operations) previously, but their experiences ranged from ``terrible'' to ``fantastic''.  All of the participants said they had experienced limitations when using their local computer for their research, with problems including lack of power, memory and storage.

\subsection{User experiences}
Typical visualisation tasks that astronomers might encounter in a data analysis workflow were presented using the local and virtual desktops.   Due to the limited time available with each participant - around 35 to 45 minutes - it was not possible to explore a wide range of applications, or to delve too deeply into the applications chosen.  We chose one task that was not overly demanding of the GPU for computation, which simulated basic desktop operations, and one where graphics card performance would be paramount.  

In a 2D environment, where no GPU acceleration is required, the task was a simple image alignment.  The focus on a specific task, rather than free exploration of the environment, provided a more objective method to evaluate whether or not the environment itself impacted on the completion of the task positively or negatively, as opposed to whether the environment was enjoyable or not.  Completing the tasks required the use of application windows, menus, keyboard commands and a mouse.  The participants were asked to inspect two FITS images \citep{abolfathi_fourteenth_2017} with DS9\footnote{\url{http://ds9.si.edu/site/Home.html}}, and align them with IRAF\footnote{\url{http://iraf.noao.edu/}}.  The participants were guided through each step of the task, and then asked to repeat the same task three times, once via a local desktop and either once or twice in a cloud environment.

\begin{figure*}
	\includegraphics[width=7in]{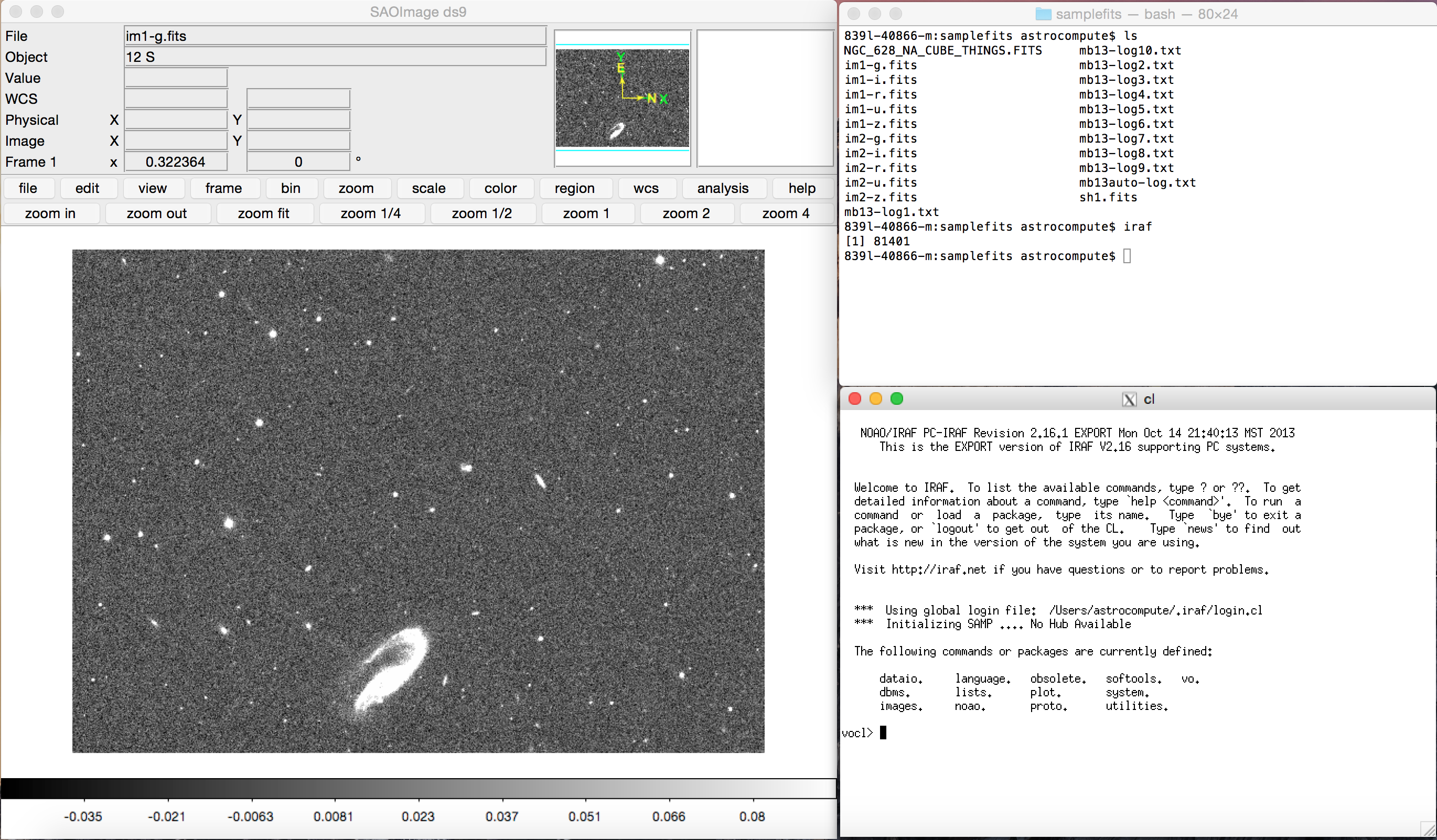}
    \caption{The layout of the display for the 2D image alignment task. The terminal window (upper right) was running before the participant began the investigation.  SAOImage DS9 (left) was launched from a desktop icon and IRAF (lower right) was started from the command line.}
    \label{fig:2dscreenshot}
\end{figure*}

Many astronomical applications require advanced 3D graphics. While it is a relatively simple matter to monitor a system's performance metrics, this does not necessarily coincide with the perceptions of an astronomer.  To determine if such a correlation exists, a GPU-accelerated 3D application called Shwirl \citep{vohl_shwirl:_2017} was used to monitor how the environment performed under increasing load, while the participants' perceptions were also recorded.  Shwirl uses graphics shaders operating on the GPU to perform interactive, real-time volume rendering of 3D spectral data cubes.   For this task, a spectral data cube\footnote{NGC628 from The H{\sc i} Nearby Galaxy Survey \citep[THINGS;][]{walter_things:_2008}, data from \url{http://www.mpia.de/THINGS/Data.html}} was loaded into Shwirl and adjustments were made to the volume rendering, mimicking steps in a workflow that might occur in visualisation and analysis of a spectral data cube.  After each change was performed, the participant was asked to provide their perception of the performance of the system at that time.

See Appendix \ref{lbl:apndx-exp-procedure} for a more detailed description of steps in the user testing procedure.

\subsection{Setup}
\label{lbl:setup}
Each participant was presented with a laptop computer with a standard mouse (to avoid possible issues with the use of a trackpad).  The starting environment was already loaded, with half the cohort seeing a local desktop first, and the rest seeing a cloud desktop first.  The cloud desktops were used in full-screen mode and minimised when not required.  Switching between desktops was performed by one of the investigators.

In all cases, to avoid having to train the participants in how to create a virtual machine or use the cloud, the environments were setup in advance, and each was preconfigured in such a way as to avoid the need for the participant to ``learn'' their way around.  Shortcuts were placed on the desktops for the applications, and terminal windows were already running.  Participants were presented with a standard desktop environment that closely resembled what they are most likely already used to.

For the NRC virtual machine, a volume mounted disk access issue caused significant delays to the loading of DS9, IRAF and Shwirl, but only for the first time they were run.  Subsequent executions did not exhibit the problem.  This is a known issue with the GPU nodes of the Melbourne Node of the NRC. To reduce the impact, the NRC virtual machine environment was prepared in advance by doing a first run of each application before the start of the user testing.  Hence, the participant experienced the cached version of the application, which closely matched the other environments. As this technical issue does not impact all NRC virtual machines or applications, it was decided this was the fairest way to compare the environments.

Every effort was made to ensure that the experience of the local and virtual hosted desktops presented to the participants was the same, and that the same set of tasks was completed.  Unfortunately, due to time restrictions imposed by working with volunteers, three of the 20 participants were only able to complete one local and one VHD version of the Shwirl task.  Additionally, a log file was not recorded for one participant while completing the Shwirl task on the AWS virtual machine.


\subsection{The 2D image alignment task}
\label{lbl:2dimageconditions}

The 2D image alignment activity was undertaken first.  The participant was provided with a sheet of paper with explicit instructions on how to proceed, including the precise IRAF commands and offsets needed to align the images. The process was timed to ensure the participant remained focused on completing the task, and to provide a means of comparing the change in performance with subsequent runs.   

\begin{table*}
    \centering
    \caption{Shwirl provides a number of modules that allow the adjustment of the loaded volume to improve visual understanding.}
    \label{tab:shwirl}
    \small
    \begin{tabular} { p{9.5em} p{18em} p{8cm} }
        \hline
        Module & Purpose & Impact on GPU\\ 
        \hline
        Camera and transforms & Adjust the field of view and scale the loaded volume in X, Y and/or Z direction & Scaling the volume in the Z direction significantly increases the computation load on the GPU                 
        \\ \hline
        Colour & Apply a colour scheme to the loaded volume & No additional calculations are required by the GPU, so impact is negligible \\ \hline
        Filter & Filter the volume to eliminate data above and below set values & No additional calculations are required by the GPU, so impact is negligible \\ \hline
        Smooth & Apply smoothing to the volume data by averaging between neighbouring points & As smoothness is increased, the range of neighbouring points increases and requires increasing number of calculations by the GPU \\ \hline
    \end{tabular}
\end{table*}

\begin{figure}
	\includegraphics[width=\columnwidth]{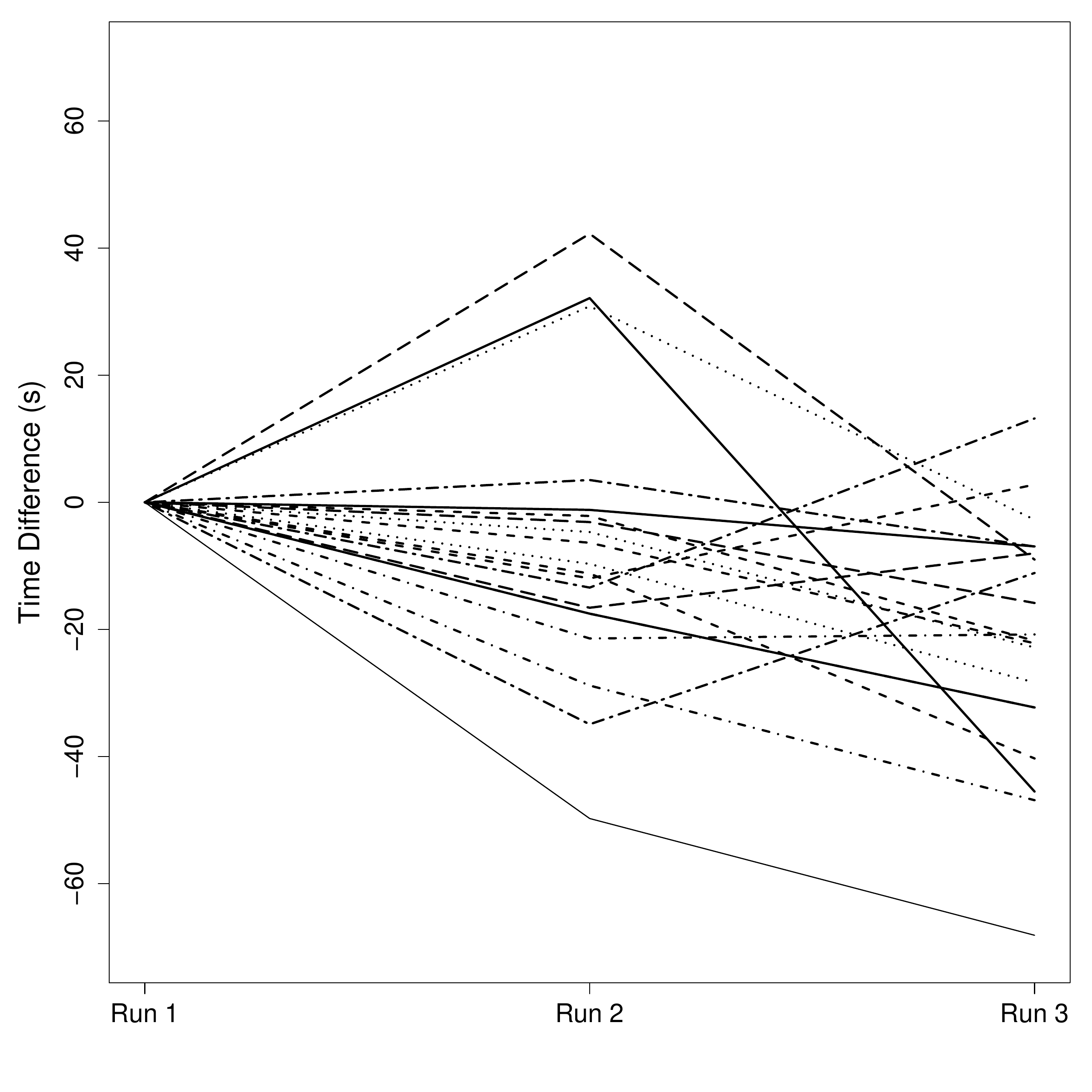}
    \caption{This graph shows the timing change from the first timed run for each participant. All but three participants showed an overall improvement, a few experienced minor problems, which caused delays.  However, these delays were caused by minor mistakes made by the participants, and not the environments.}
    \label{fig:training_effect}
\end{figure}

To ensure all participants were familiar with the instructions, the operator guided the participant through every step without the timer running.  Once the participant had completed this initial pass, they were then instructed that they would be timed for the next three passes.  After some of the participants made minor mistakes that resulted in noticeable delays, it became clear that it was necessary for the operator to intervene if it was apparent that an error was about to occur, such as missing a step, that would result in a significant loss of time.  This was deemed acceptable because our purpose was not to evaluate the participants' ability to learn a task. 
See Figure \ref{fig:2dscreenshot} for the screen layout during this activity.

The participants were timed by one of the investigators while completing the task, but were encouraged to simply ``follow the instructions'' without worrying about the timing. By repeating the task, participants generally improved their time, which suggests that they were learning while working.  Such learning effects are expected in task-based user studies.  Varying the order in which participants were presented with a local or cloud desktop was necessary to determine if the environment itself impacted on this learning process.  Timing the process encouraged the participants to focus on performing the task itself rather than whether the environment lived up to their expectations.


For the 2D image alignment task, the cloud and local environments performed equally well.  Figure \ref{fig:training_effect} shows the individual time changes based on each participants' initial time, with most participants improving their performance each time.  In each case where a participant saw a decrease in performance (corresponding to an increase in time taken to complete the task), a simple error such as clicking the wrong button or misreading an instruction, was identified as the cause of the problem.  In only one case did a participant attribute a loss of performance to the environment, and in that case, it was the lack of familiarity with the local desktop operating system (MacOS X) that was identified as being the problem.  One other participant encountered a minor issue where they accidentally switched out of the virtual machine client.  In that instance, because it was the first time trial, and no steps had been completed, the task and timer were restarted.  The order of presentation of the environments did not appear to have any impact on the performance or the participants.

Based on user testing, we conclude that both the local and virtual hosted desktops were equally suited for tasks of this nature, especially before significant demand was placed on the graphics systems.  Participants reported very positive experiences with the VHDs, and many expressed surprise at the level of performance and ease of use for the cloud environments.


\subsection{The 3D spectral cube rotation task}
\label{lbl:3dcuberotation}

\begin{figure*}
	\includegraphics[width=7in]{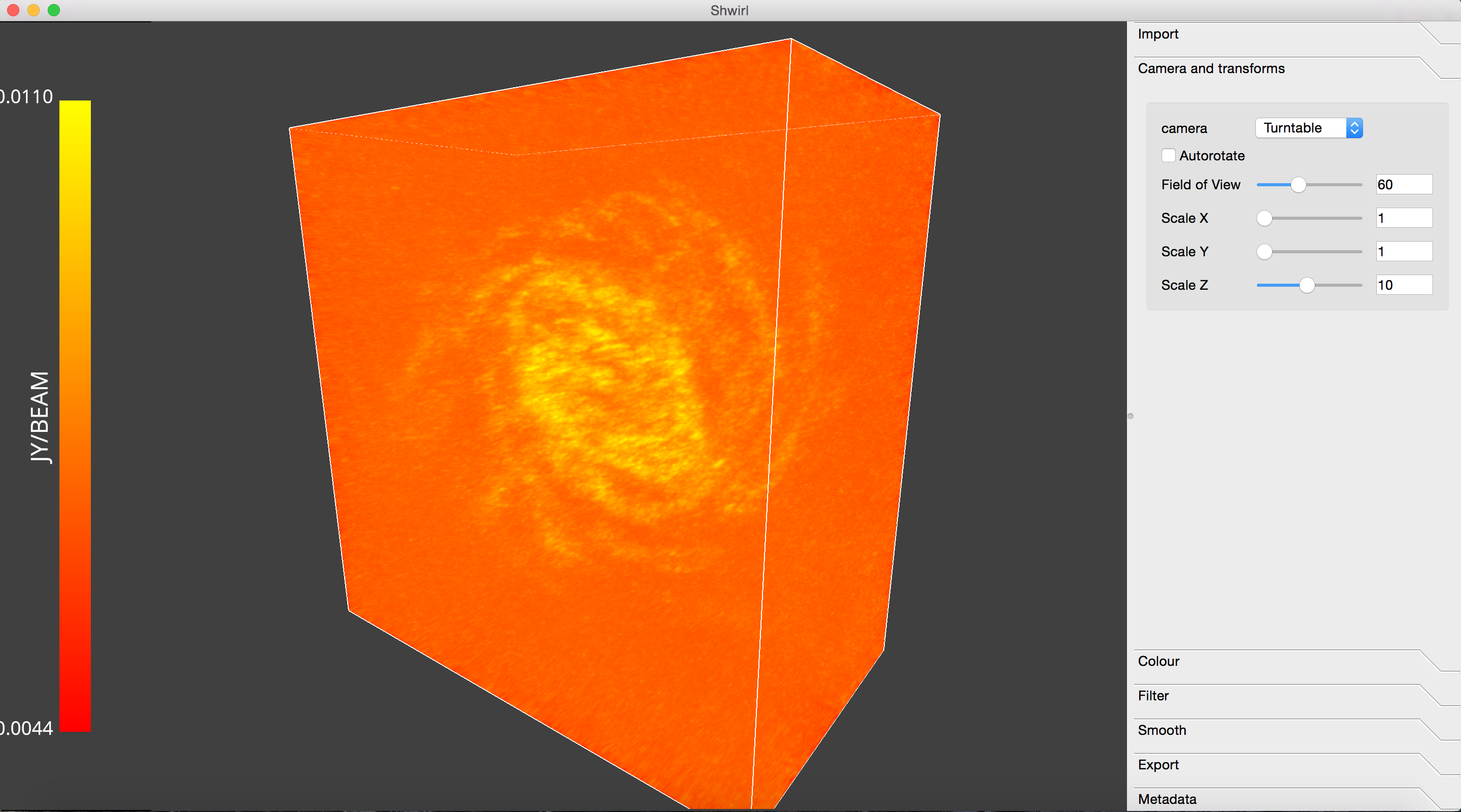}
    \caption{The Shwirl application was launched from the command line. The tabs on the right side of the application are the Shwirl modules (see Table \ref{tab:shwirl}), and correspond to the tasks for each step of the 3D activity.  This image shows the spectral cube with a scaling of 10 in the Z direction.}
    \label{fig:3dscreenshot}
\end{figure*}

The 3D spectral cube rotation activity was not timed by hand because the Shwirl software generates a log of the frame rate once per second, linked to the corresponding states of the Shwirl options.  This provided a means to compare the system performance with the participants' perception scores.  See Figure \ref{fig:3dscreenshot} for the screen layout during this activity.

\begin{figure*}
    \includegraphics[width=6in]{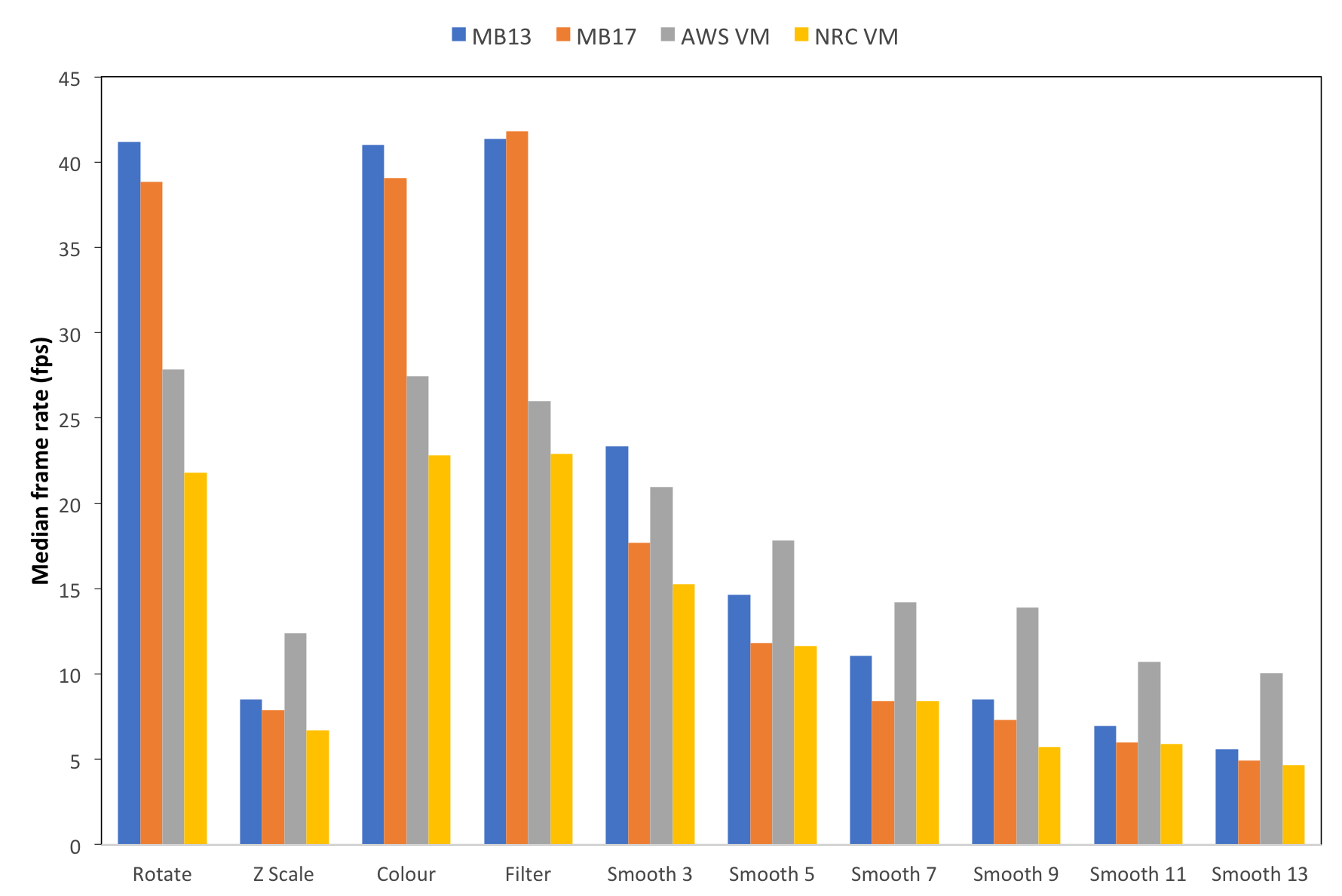}
    \caption{Using the auto rotate feature in Shwirl, each of the steps in the 3D component of the task were completed to establish a benchmark for the platforms.  The Rotate step shows the environments without additional computation load, while the Z-Scale shows an increased load.  These steps were included to give the participants the opportunity to calibrate their perception.  The steps are included here as they correspond to the steps undertaken by the participants.}
    \label{fig:fps_auto}
\end{figure*}

\begin{figure*}
    \includegraphics[width=6in]{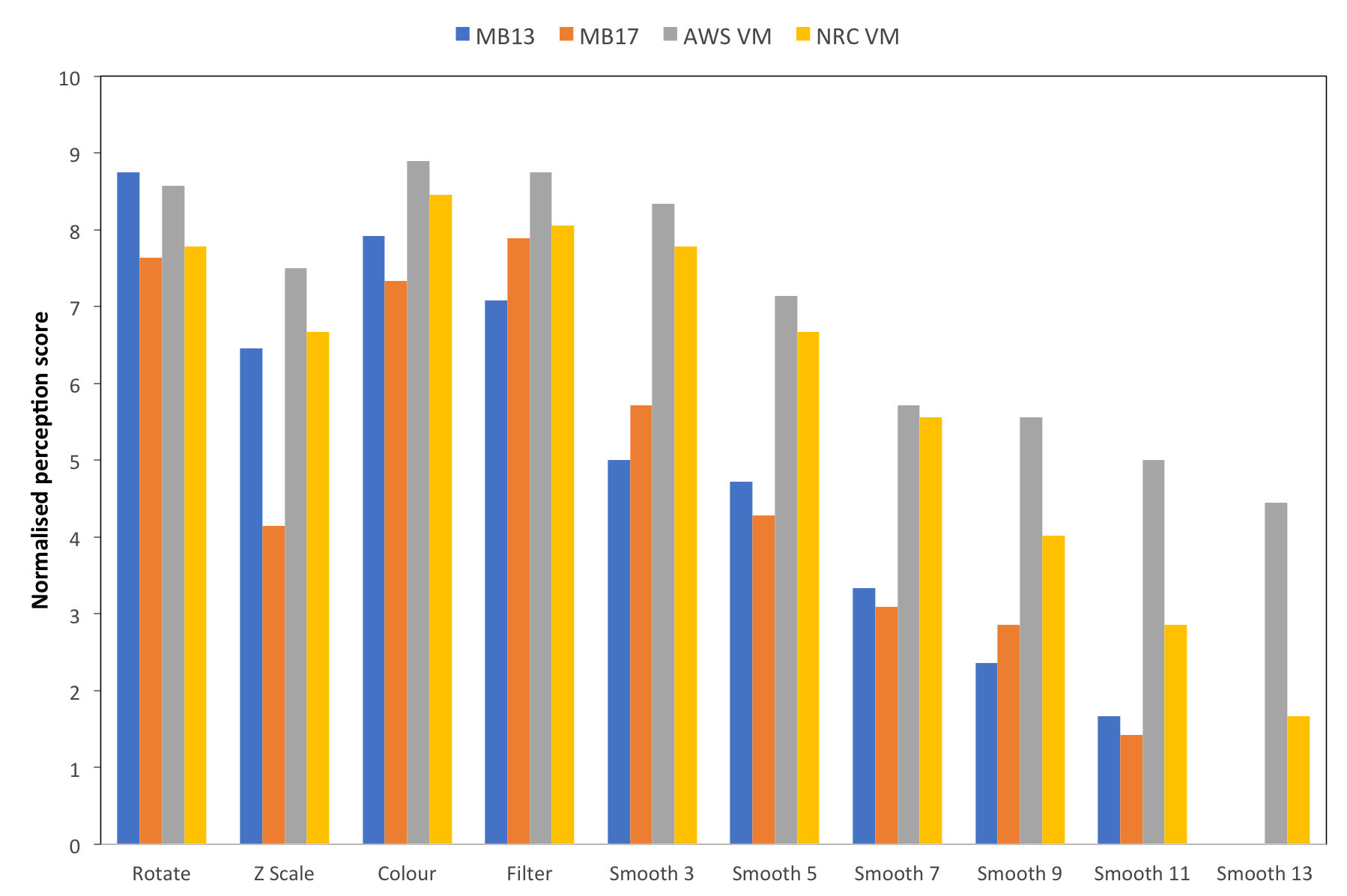}
    \caption{After each step of the 3D tasks, the participant scored the environment between 0 and 10.  The normalised median perception scores are shown for each environment and for each stage of the 3D component of the investigation.  The overall trend follows the one seen in \ref{fig:fps_auto}. Each of the participants had access to both the cloud environments and only one local laptop, so the above scores are each drawn from 10 $\times$ MB13, 10 $\times$ MB17, 18 $\times$ AWS virtual machine, and 18 $\times$ NRC virtual machine results.}
    \label{fig:perception}
\end{figure*}

\begin{figure}
	\includegraphics[width=\columnwidth]{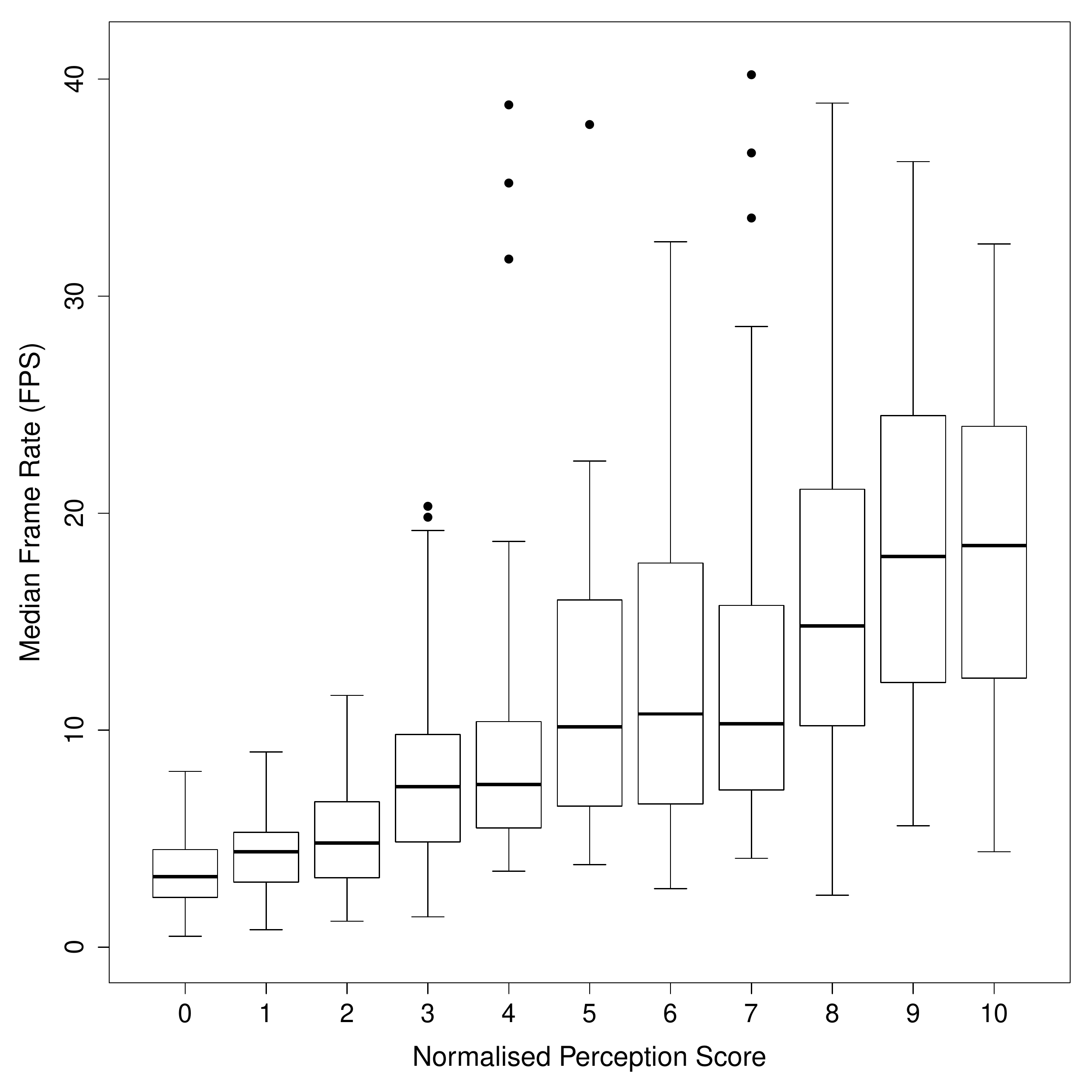}
    \caption{The normalised median perception scores versus the median frame rates shows that in general, participants' increasing perception scores corresponded with the increasing frame rates. However, there is also considerable variation, and sometimes high frame rates correspond to low scores, and low frame rates can correspond to high scores. This reaffirms the idea that frame rates are not the only factor participants consider when evaluating an environment.  For each Score, the whiskers extend to the data value that is no more than twice the interquartile range from the box.}
    \label{fig:perception_vs_fps}
\end{figure}

As the participant completed each section, they gave the environment a score out of 10, where 10 was identified as being ``as good as they could possibly want'' and 0 being ``something they would never willingly use again''.  As this measure is very subjective, the responses were normalised for each participant, such that the highest score they gave for all three environments became a 10, and the lowest score they gave became a 0.  This provided a direct comparison of the participants' perception of their experiences of the two or three environments encountered. These results were then compared with the system performance for each stage of the 3D tasks.

The participants' responses were recorded for each step and environment, along with a Shwirl output log for each.  This way, each log could be associated with the corresponding perception scores. Table \ref{tab:shwirl} shows the modules used during the task and how they are used to test the GPU in the environment.  More detailed descriptions of the modules can be found in \citet{vohl_shwirl:_2017}.



Using Shwirl's auto-rotate feature, the same steps to be used by the participants were applied to create a baseline for the investigation.  Figure \ref{fig:fps_auto} shows the median frame rate for each environment for each stage of this task.  This graph shows that the GPUs in the local laptops initially performed much better than the GPUs in the cloud environments.  However, as the load was increased on the GPU, such as in the Z-Scale and the smoothing steps, the performance of the local GPUs dropped more than the cloud environments.  Once smoothing was applied at increasing levels, the difference in the performance became almost negligible, with the AWS virtual machine decreasing at a slower rate than the others.

\subsubsection{User perceptions of computing environments}
Participants naturally have different approaches to assessing an environment, so it was necessary to provide some guidance to ensure some similarity in the assessment.  The very first time a participant was asked to provide a ranking for an environment, they typically opted for numbers around 7 or 8, as these reflect the generally positive experience, without having over committed.  This approach gave the participant room to give subsequent environments a higher score if they performed better, or lower if they performed worse.  

The first two tasks were intended to give the participant an opportunity to calibrate their perception.  The first task provided an unaffected interaction with the loaded volume, while the second task placed considerable load on the GPU.
Despite this calibration step, it was still necessary to normalise the responses.

Figure \ref{fig:perception} shows the corresponding stages of the 3D spectral cube rotation as median perception scores for each environment.  The overall trend follows the one seen in Figure \ref{fig:fps_auto}:
\begin{enumerate}
    \item a relatively high initial score;
    \item a drop for the Z-Scale step;
    \item a return to the previous levels for the Colour and Filter steps;
    \item a steady decline for the smoothing steps.
\end{enumerate}

It is interesting that despite the high frame rates for the local laptops in the Rotate, Colour and Filter steps, the perception scores for all four environments are very close.  This suggests that the usefulness of an environment is acceptable above a certain frame rate, and that additional performance does not necessarily correspond to a better experience.

\subsubsection{Performance}
Manipulation of a large spectral cube places a significant load on a GPU. Most laptop GPUs are designed to provide an optimal GUI experience, and are not designed for these types of  workloads. 

Increasing the Z-Scale factor to 10 had a huge impact on the GPU performance, with all environments dropping significantly.  This resulted in an expected drop in perception score, though the MB17 showed a much bigger drop than the others. When the Z-Scale was returned to 1, and a colour filter was applied, the frame rates returned to the previous levels, as did the perception scores, with a slight improvement for the cloud environments over the local environments.

Applying the Colour or Filter option does little to affect the GPU performance and this is reflected in frame rates as seen in Figure \ref{fig:fps_auto}.  As expected, the perception scores also remained relatively unchanged, still with a slight favouring of the cloud environments.

\begin{figure*}
    \includegraphics[width=\columnwidth]{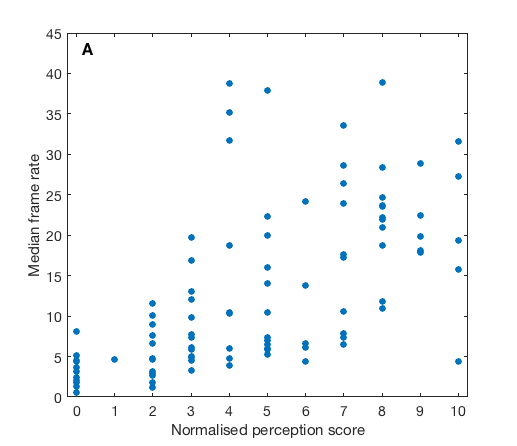}
	\includegraphics[width=\columnwidth]{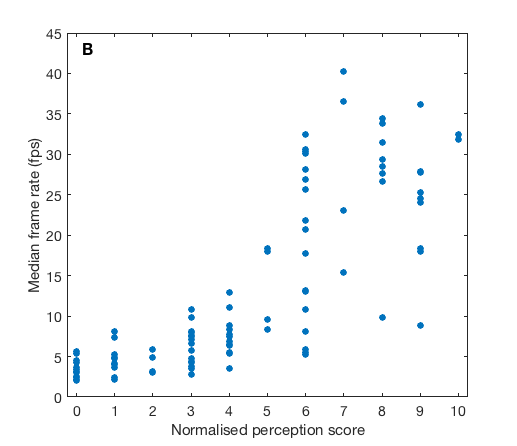}
	\includegraphics[width=\columnwidth]{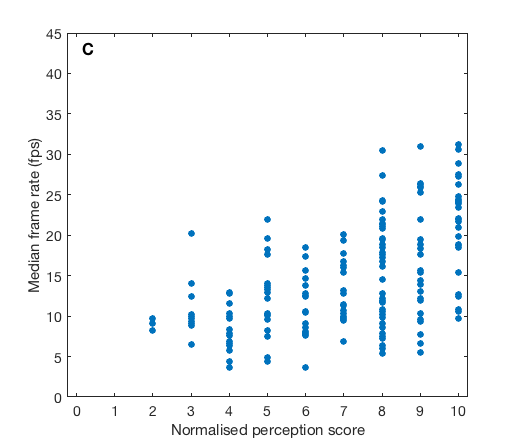}
	\includegraphics[width=\columnwidth]{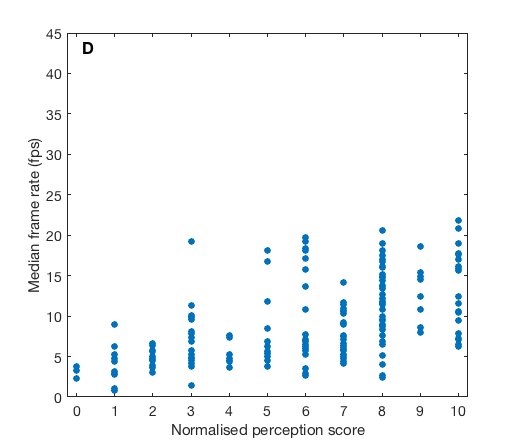}

    \caption{The normalised median perception scores versus the median frame rates for the individual environments shows greater variation in the local laptops (A) MB13 and (B) MB17 than the cloud environments (C) AWS virtual machine, and (D) NRC virtual machine. This suggests that the variability in the frame rate is linked to the user experience.}
    \label{fig:perception_vs_fps_environment}
\end{figure*}

As before, when a greater load is placed on the GPU, in this case applying a Smoothing algorithm, the local computer GPU frame rate drops markedly.  Oddly, MB17 dropped more than the MB13, which suggests that the GPU in the MB13 is better suited to this sort of workload than the one in the MB17 (see Section \ref{lbl:systems_specs} for more details).  Yet at smoothing value of 3, the cloud virtual machines' perception scores are now markedly higher than the local computers.

As the smoothing value is steadily increased to 13, we see a steady drop in all perception scores, though a clear separation between the local laptops and the cloud environments is apparent.  By smoothing value of 5, the AWS virtual machine outperforms the other environments and continues to do so for the rest of the task.  For the NRC virtual machine, even though its frame rate for the higher smoothing values are almost the same as the local computers, it maintains a higher perception score until the end of the task.

\subsubsection{Perception versus frame rate}
Figure \ref{fig:perception_vs_fps} shows that while there is a correlation between the normalised perception score and the median frame rate, a great deal of variation is still present. Figures \ref{fig:perception_vs_fps_environment} A, B, C and D show that the variation of frame rate on the local environments is considerably greater than the cloud environments, and likely is the most significant factor for determining the perception score.  The bigger the drop in the frame rate experienced, the lower the perception score, regardless of the highest frame rate experience.  This is why low perception scores are observed even for the highest of frame rates.  We also see some surprisingly high perception scores for low frame rates, suggesting that for some participants, the frame rate itself did not determine their experience of the environment for a particular task.  That is, a low frame rate can still provide a satisfactory experience for some researchers, depending on their expectations and needs.

\subsection{Post-task user reflections}
\label{lbl:postexpsurvey}
\begin{figure}
	\includegraphics[width=\columnwidth]{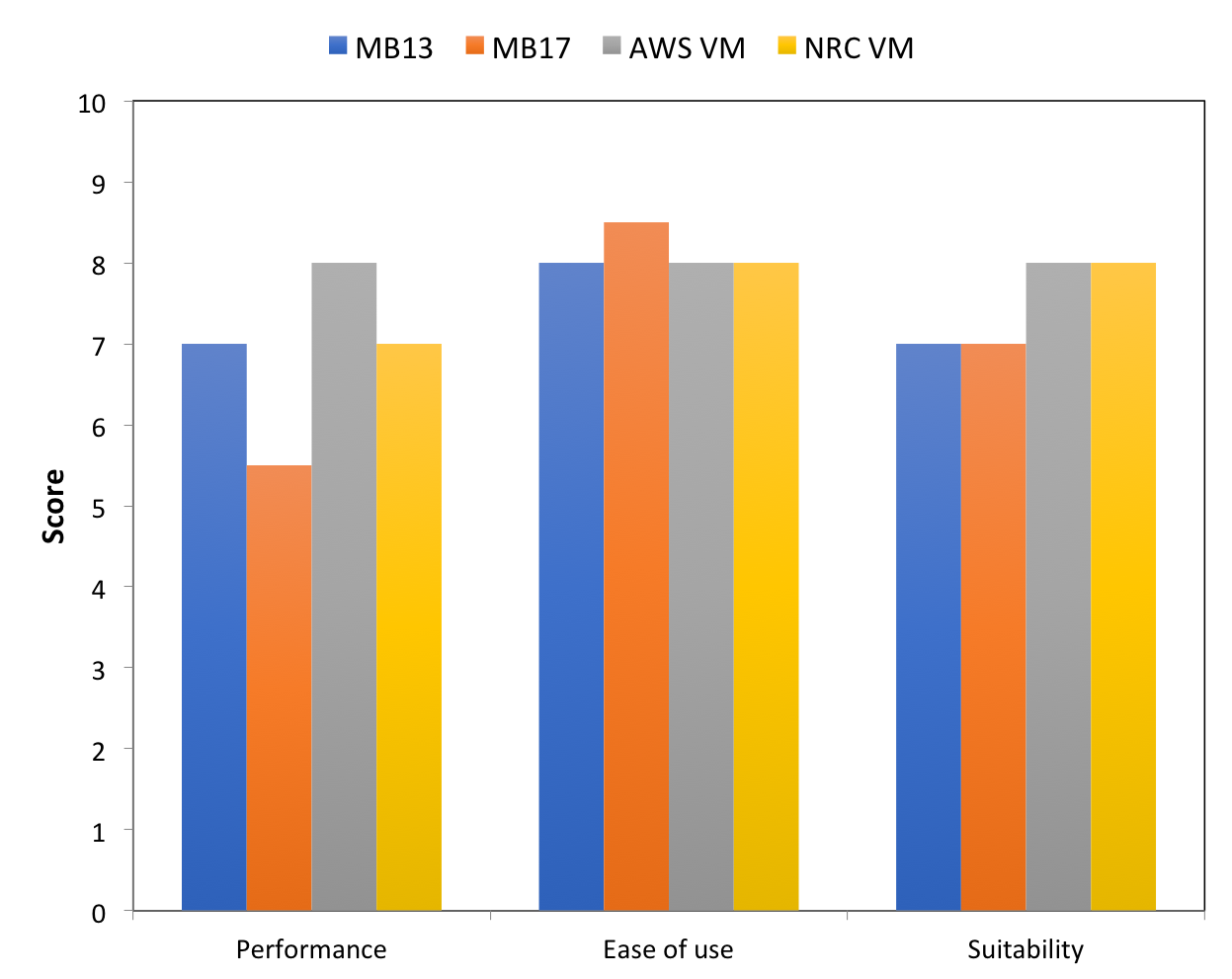}
    \caption{Having completed the tasks, the participants were asked to rate (out of 10) the NRC and AWS virtual machines and the laptop they used for Performance, Ease of use, and Suitability.  The values shown are the median response scores for the whole cohort, with 10 responses for MB13 and MB17, and 18 responses for AWS virtual machine and NRC virtual machine. The performance of the MB17 was rated lowest, despite showing the best performance using a standard benchmark (see Figure \ref{fig:UnigineBenchmarks}). Overall, the cloud environments were seen as being equally performant, easy to use, and suitable for the tasks undertaken.}
    \label{fig:survey}
\end{figure}

After completing the hands-on component of the investigation, the participants were asked to reflect on their experience with either the 2D or the 3D task.  They were asked again to rate the environments they had used from 0 to 10, but this time they were focusing on performance, ease of use, and suitability for the tasks conducted.  Four of the participants chose to focus on the 2D experience, while 16 chose the 3D experience.  This split might have occurred as the 3D experience was more readily recalled because it more recent, or it may have been chosen because it was perceived as more enjoyable.  Since the purpose of the post-investigation survey was to focus the participant on the functional purpose of the environments, and to be able to meaningful comment on the viability of the VHDs as a replacement for a local desktop computer, this split does not affect the overall outcome.  The survey results are shown in Figure \ref{fig:survey}.

\begin{itemize}
    \item \textbf {Performance:} 17 participants had previous experience with X11 Forwarding or VNC, and they generally found its performance adequate for the tasks they required it for.  However, after having completed the user tests, these participants found the cloud environments matched or outperformed both the local environments and their previous experiences.  While the best frame rate was recorded for the MB17, it also received the lowest score for performance.  Figures \ref{fig:perception_vs_fps_environment}A and B, reveal greater variation in the median frame rates than the cloud environments shown in C and D, which influences participants' perceptions and hence their overall impression of the environments.  
    \item \textbf {Ease of use:} While MB17 received the lowest Performance score in the post-investigation survey, it received the highest score for Ease of use.  While not specifically recorded, most participants were quite comfortable with the MacBook Pro laptops and scored them positively.  Overall, the four environments evaluated in the investigation were considered easy to use, which suggests that once familiarity is gained in a cloud environment, there is little difference between operating a virtual machine and a physical computer.
    \item \textbf {Suitability:} This score was intended to give the participants the opportunity to summarise their experience.  As Figure \ref{fig:survey} shows, the participants generally found that while the local environments were quite acceptable, they were exceeded by the cloud environments.  Interestingly, despite the AWS virtual machine showing a clear lead in the perception scores for the 3D tasks, it was seen as being just as suitable as the NRC virtual machine.
\end{itemize}

\begin{figure}
	\includegraphics[width=\columnwidth]{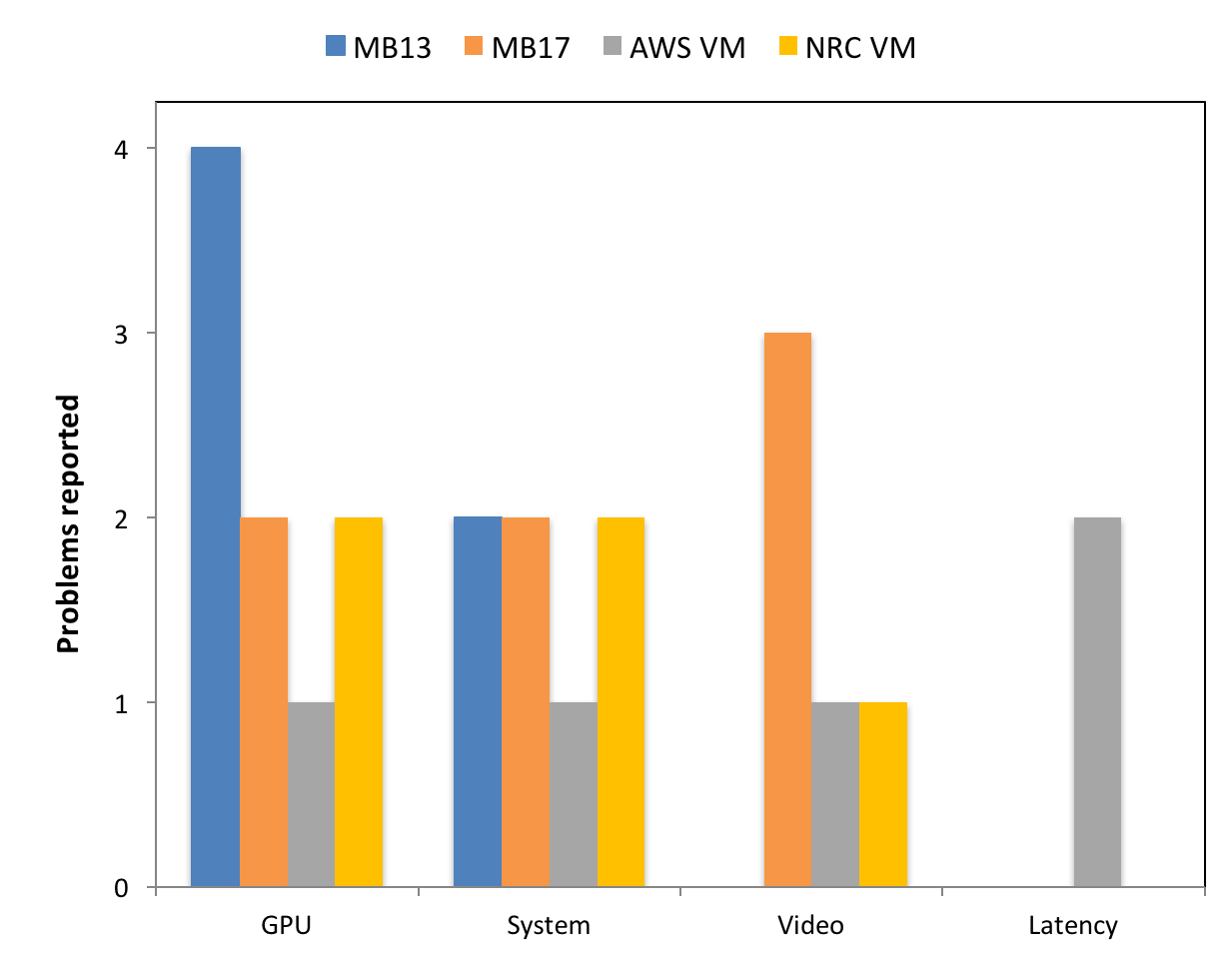}
    \caption{During the post investigation survey, participants identified issues with the environments.  These can be categorised into four themes, relating to the GPU, the System, the Video display and Latency.  The MB13 GPU was the most complained about, followed by the video tearing on the MB17. Latency was only observed for the AWS VM. In general, participants experienced fewer problems with the cloud environments.}
    \label{fig:problems}
\end{figure}

During the post-investigation interview, the participants were asked to reflect on the problems they experienced with the different environments they used.  Figure \ref{fig:problems} shows the frequency of these issues, which are categorised into four themes:

\begin{itemize}
    \item \textbf{GPU:} The GPU performance was considered the main problem for the completion of the task
    \item \textbf{System:} The system experienced a momentary freeze, crashed, was unfamiliar to the participant, slow to load the application, or slow to load data
    \item \textbf{Video:} Most commonly experienced as video tearing, where a mismatch in the GPU rendering frequency and the screen display refresh caused a momentary splitting of the image. Also where the colours displayed were not as expected
    \item \textbf{Latency:} the only noticeable demonstration of this issue was when the DS9 application window was moved, but tracked slightly slower across the screen than the mouse
\end{itemize}

Notably, the MB13 received the most complaints about the GPU performance, however it was the only one not to receive comments about video tearing.  Video tearing was most apparent on the MB17, while the AWS virtual machine was the only environment where the latency was an issue.  


\section{Discussion}
\label{lbl:discussion}


To be a viable replacement or adjunct to a desktop machine, a VHD needs to support research workflows in a seamless
way.  The user experience -- compared to a physical desktop -- must be maintained or even exceeded. Astronomers are more likely to accept
and adopt the use of a fully remote desktop if the interaction speeds (e.g. from movement of
mouse on the desktop to movement of the mouse cursor on screen) are no worse than those
currently achieved on a local desktop.

The key outcomes of our user experience testing demonstrate the following:
\begin{itemize}
    \item VHDs are as easy to use as a standard desktop;
    \item A correctly resourced and configured VHD provides a suitable environment to run typical astronomy software;
    \item A correctly resourced and configured VHD can provide a better user experience than a local laptop computer; and
    \item VHDs can provide a viable desktop alternative for astronomers.
\end{itemize}

However, the tasks did not touch upon some other elements of VHDs that are part of the cloud experience.  For example, creating, configuring and maintaining a virtual machine is a non-trivial task and requires a reasonable level of technical skill.  Factoring this aspect into the experience might have had a negative impact on the impressions of the participants, and therefore it might be argued that this should have been included.  Yet cloud specific impacts need to be measured against local computer impacts.  

It is also important to consider whether pre-built cloud images can mitigate much of the challenge associated with the creating, configuring and maintaining a virtual machine.  Further, a managed service can eliminate the need for an astronomer to manage the VHD entirely.  Because of the complexity surrounding the establishment of a VHD in comparison to a local computer, the investigation was deliberately limited to focus purely on the operational use of the environments.

While all participants agreed that there was considerable potential in the use of VHDs in their workplace, some skepticism remained.  The idea of committing to a cloud-based service continues to be a source of concern for many participants, and reflects the wider community attitudes.  However, there are many elements to consider when choosing a suitable environment.

\subsection{Pricing comparison}

\begin{table*}
	\centering
	\caption{This table shows the basic pricing for each environment, and the life expectancy of the hardware purchased. The AWS On-demand and Spot pricing are included separately. An ASUS i7 laptop is included for comparison, though one was not used during the study.  VM = Virtual Machine. UoM = University of Melbourne.  Prices are shown in Australian dollars, noting that there are also country-specific variations for hardware purchases.}
	\small
	\begin{tabular} { p{9.75em} p{14em} p{4cm} p{5cm} }    
		\hline
		Environment & Approximate Price & Life expectancy & Comment\\
		\hline
		MB13 & \$3,000 (2013) & 4 years & Best option available in UoM standard catalogue\\
		MB17 & \$4,500 (2017) & 4 years & Best option available in UoM standard catalogue\\
		NRC VM & \$200 per core per year (2017) & New hardware added each year & Limited GPU options\\
		AWS VM (On-demand) & \$0.90 per hour (2017) & New hardware added each year & Additional costs for storage and data ingress/egress\\
		AWS VM (Spot) & \$0.34 per hour (as at 2017-11-15) & New hardware added each year & VMs are stopped at bid threshold\\
		ASUS i7 & \$2,100 (2018) & 4 years & Comparable system to the MB17, from the Amazon online catalogue (Australia)\\
		\hline
	\end{tabular}
	\label{tab:pricemodels}
\end{table*}

The initial outlay for a computer of reasonable power is something that many researchers and research departments take for granted as they prepare their annual expenditure projections.  For a typical PhD research student, an estimate of the likely computation power they will require is made at the start of their research journey. However, it may take several months before they reach a point of actually needing that power.  Until that point, they are financially over-committed on the purchase.  Having started to make reasonable use of the computer, they may well find that soon the demands of their research outgrows their local machine, and they need to move to remote high-performance computing facilities.  Once again, they are over-committed on the purchase of the computer.

To determine how powerful a computer should be to meet a researcher's need, the most common approach is to estimate the peak workload expected, and then buy a machine that most closely achieves that requirement within the available budget envelope.  However, unless tested directly with the workload intended, simple benchmarks and specification comparisons may not provide an adequate indication of the machine's suitability.  Combined with the fact that the machine may only be used at that peak for a fraction of its lifetime, a financial over-commitment is likely.

Alternatively, a more modest purchase paired with a suitable virtual machine from a cloud provider could provide a far more financially responsible option.  Furthermore, an out of warranty machine may still be able to perform sufficiently well as to provide connectivity to a virtual hosted desktop for several additional years.  A local computing device and a suitable network to access the virtual machine is still required, but the local device does not need more power than is required for that function.

A service like the Nectar Research Cloud is not directly charged to the researchers for research supported by national funding, and several participating institutions also provide additional resources internally to their own researchers.
Commercial service providers like AWS are fee-for-service and are an excellent way to explore options without committing significant financial resources.  See Table \ref{tab:pricemodels} for a pricing table.  

More extensive testing with a larger range of laptop and desktop computers, as well as other virtual machine variants and locations, would have been ideal. However limitations on participant time and numbers, as well as the lack of financial resources to purchase additional equipment beyond the laptops included in the study, made this impossible. Furthermore, as \citet{lam_empirical_2012} indicates, the immediate subjective feedback from user experience is more important than the range of technical specifications in this case. While a Linux laptop was not able to be purchased for the study, a model with similar specifications to the MB17 (an ASUS i7-7700HQ\footnote{Specification: 16GB DDR4 RAM, 256GB SSD drive, 15.6inch Ultra HD display, NVIDIA GTX1050 (4G RAM), Microsoft Windows 10.  Purchase price of $\sim$AUD\$2,100.00 was correct at 18/03/2018} with a more performant graphics card than the MB17) is included in Table \ref{tab:pricemodels} for the purpose of comparison. The model ships with Microsoft Windows 10 but could be reinstalled with the same Linux Ubuntu operating system and applications used in this study. 

The pricing comparisons in Table \ref{tab:pricecomparison} show the costs of laptops and the cloud environments under different usage models.  Hardware purchased, either as laptops MB13 or MB17, or as part of the NRC provides, provide the best value for money for the maximum use scenario, where the computer is used at capacity for its useful life.  However, other than high performance compute clusters, this is an unlikely usage pattern.  The high use scenario, where the device is used 6 hours per day, 5 days per week, 42 weeks per year for four years, shows a close alignment between the MB13 and the NRC VM, and again with the MB17 and the AWS VM.  However, based on the experiences reported by the participants, the cloud environments would still provide the better value for money. Finally, the low use scenario consisting of 2 hours per day, 3 days per week, 32 weeks per year for four years, shows a clear price advantage for the AWS VM, costing less than \$700.  As above, the ASUS i7 is included in Table \ref{tab:pricecomparison} for reference.

The above discussion focuses on the On-demand pricing for the AWS VM, and further cost reduction can be made by using Spot pricing, though additional care needs to be taken to monitor the resource availability.  It is also important to note that the cloud pricing for both NRC VM and AWS VM include all costs, while the laptop costs include only the initial purchase price.

\begin{table*}
	\centering
	\caption{Direct pricing comparisons can be difficult due to the difference in usage and service models.  This table shows the price comparison for a maximum use model (i.e. 24 hours per day, 7 days per week, 52 weeks per year, 4 years), high use (6 hours per day, 5 days per week, 42 weeks per year, 4 years), and low use (2 hours per day, 3 days per week, 32 weeks per year, 4 years). * indicates the fixed value.  Non-starred values are calculated based on the fixed value and the maximum, high and low usage configurations. NRC virtual machine priced based on estimated \$200 per core per year.  An ASUS i7 laptop is included for comparison, though one was not used during the study. Prices are shown in Australia dollars.}
	\begin{tabular}{lcccr} 
		\hline
		Environment & Approximate Price & \$ per hour (maximum) & \$ per hour (high) & \$ per hour (low)\\
		\hline
		MB13 & \$3,000* & \$0.09 & \$0.60 & \$3.91\\
		MB17 & \$4,500* & \$0.13 & \$0.90 & \$5.86\\
		NRC VM & \$3,200* & \$0.09 & \$0.64 & \$4.17\\
		AWS VM (maximum, On-demand) & \$31,379.71 & \$0.90* & \$0.90* & \$0.90*\\
		AWS VM (high, On-demand) & \$4,525.92 & \$0.90* & \$0.90* & \$0.90*\\
		AWS VM (low, On-demand) & \$689.66 & \$0.90* & \$0.90* & \$0.90*\\
		AWS VM (maximum, Spot) & \$11723.71 & \$0.34* & \$0.34* & \$0.34*\\
		AWS VM (high, Spot) & \$1,690.92 & \$0.34* & \$0.34* & \$0.34*\\
		AWS VM (low, Spot) & \$257.66 & \$0.34* & \$0.34* & \$0.34*\\
		ASUS i7 & \$2,100.00* & \$0.06 & \$0.42 & \$2.73\\
		\hline
	\end{tabular}
	\label{tab:pricecomparison}
\end{table*}

\subsection{Risks}
When considering the viability of a VHD as a replacement or augmentation of an existing desktop computer, it is important to consider the risks.  As part of the post-experience evaluation, participants reported on their concerns. Possible solutions are available to each of the risks using either direct mitigation or user education.

\subsubsection{Network availability}
The concern raised most often by the participants when considering the viability of a VHD was the availability of a suitable network.  The network needs to provide sufficient bandwidth and low latency, but must also remain consistent.  This concern reflects the changing way that researchers work, with portable devices and working in a variety of conditions with a variety of networks now common.  Most research institutions now recognise the importance of providing a highly stable network.

However, while campus networks might be stable, the local caf\'e or airport wifi may not be.  Other networks such as ADSL or cable connections in the home, or 4G mobile networks, are more subject to congestion than campus networks, and therefore may provide a suitable network at some times, and not others.

While network connectivity is important when using VHDs, an unexpected disconnection is disruptive to workflow, VHDs are quite tolerant of this sort of event.  When a client suddenly loses connectivity, the VNC connection is dropped, but the running VHD session does not terminate.  Instead, it remains active and available for when the connection to the client can be re-established.


\subsubsection{Hidden costs}
Using a cloud service like AWS requires careful planning to ensure costs do not blow out.  While the On-demand pricing captures the essential rate for the virtual machine, it does not include things like data ingress and egress, or additional storage capacity.  While there are several tools for monitoring active resources, it is quite easy to overlook certain components, which can result in a significant bill.

For a service like the Nectar Research Cloud, the hidden costs are more significant.  For the individual researcher, the service can appear to be completely free, as they never see a bill, nor are they necessarily required to answer to their department for their usage costs.  However, institutions like Nectar, the University of Melbourne and Swinburne University of Technology invest capital and operational funds to provide the service, and are answerable for the efficacy of that expenditure.  Without a mechanism to remind researchers of the value of the service they consume, the claimed resources are often left idle.  For many such cloud services, the resource utilisation can be as low as 10\%, even as the resource allocation approaches 100\%. 

\subsubsection{Security}
As more research is conducted online, the security of online resources is of ever increasing importance to researchers and their institutions.  Most researchers are familiar with online storage, but many are not aware of the potential security implications.  This includes the possibility of hacking, ethical considerations, and the unwitting loss of intellectual property rights.

Hacking is one of the highest concerns, where unauthorised access to research data can occur.  In the case of a VHD, there is an additional risk that the VHD itself may be managed by a non-expert, who may not be as proactive in securing a system as they need to be.  A breach might also occur with a properly secured system, simply because of a bug in the connection service, or one of the many other system services, that can be exploited by a person with nefarious intent.  However, training and diligence are very effective tools for minimising the possibility of this happening.

In some cases, though unlikely in the field of astronomy, the data being processed might have ethical requirements that prevent it being stored in a public cloud, or prevent it being moved out of a geographical location.  This is a common issue for medical and biological science data.  In this case, the use of private research clouds is usually an acceptable approach.

Finally, the storing of data on a public cloud may subject the data to intellectual property laws that relate to the location of the data centre rather than the origin of the data.  Protections applicable in one country are not necessarily available in another country.  It is important that researchers are aware of where their data will be stored, and what laws have jurisdiction over their data.

\subsection{Convenience}
The convenience of using a laptop or desktop computer, without needing to connect to a VHD, was cited by one participant as being their main reason for not accepting a VHD as a viable alternative.  However, as common as this view might be, it does not take account of the risks associated with the local computer itself.  For example, a local computer can fail for many reasons, such as loss of power, breakage, or theft.  While these issues also prevent access to a VHD from that device, they do not prevent access from another available device.  In fact, reducing the dependence on a specific local device reduces the risk of mid-to-long term access to resources.  Theft from a properly secured data centre is rare, and while hardware failure is quite common, cluster-based computing, and the services which operate on top, are typically very tolerant of such failures due to their frequency.

Interestingly, the criteria defined in Section \ref{lbl:clouddesktops} are measured in comparison with the operation of a local computer, but such a local computer need not meet these criteria to be considered acceptable. One participant stated that even though they recognised that the cloud environments performed better, they would still choose a physical desktop computer over a cloud-based desktop solution, even if it did not perform as well, because it could be used in places where the network might not be adequate.

\section{Conclusions}
\label{lbl:conclusion}

As we move closer to the Square Kilometre Array's (SKA) exabyte-data era, it will be increasingly impractical to visualise data products on a local desktop with a standard display device \citep{hassan_scientific_2011}.  Not only will the data vastly exceed the local storage, memory and computation capabilities of desktop computers, but the image resolution and graphics capabilities required will force astronomers to consider large-format displays and specialised graphics processing hardware.

Today, many instruments already produce datasets that are too big to be downloaded to a local computer, so it stays in the data centre.  The compute power also sits in the data centre, close to the data.  Astronomers are happy to use remote computing as part of their workflow, but prefer to continue use desktop applications for some tasks, such as visualisation.  However, this typically means transferring data to the local computer.  As this becomes harder to do, having windowed applications forwarded to the local desktop is one solution, but the performance is limited.  

Cloud services with GPU hardware allow GPU-enabled applications to be used in a VHD environment on a remote computation platform, with just the screen updates transmitted to the local computer.  




By combining benchmarks with user experience testing, we found that VHDs provide a viable, cost-effective desktop alternative for typical astronomy applications, particularly for graphics-intensive tasks.  

While benchmarking may approximate the intended workload, only direct testing with astronomy workflows operating under load will really determine the adequacy of a compute resource. Standard benchmarking applications, Unigine Heaven and Valley, indicated that a new laptop computer was more performant than a cloud-based VHD, but user experience testing revealed that for some tasks, a VHD can provide a better solution.

Through our combined use of benchmarking and user experience testing we compared two laptops, one four years old and one new, with two cloud-based desktop environments (AWS and the Nectar Research Cloud).  We have shown that:
\begin{enumerate}
    \item For the 2D and 3D tasks,  the environments were equally simple to use.  During the investigation, it was apparent that the cloud environments operated as hoped for the tasks presented, easily matching the local computers.  As many astronomers use standard applications, and these applications function the same way across the available environments, the only differences are the environments themselves.
    \item The cloud environments provided an equally smooth experience for the required tasks as the local computers.  Figure \ref{fig:problems} shows that other than minor latency caused by the distance to a data centre (in the case of AWS virtual machine located in Sydney, Australia), the cloud environments matched or bettered the local computers. In fact, one of the likely reasons for the positive responses to the cloud environments was due to the more consistent performance shown by the cloud environments when compared to the more erratic performance (in terms of frame rate) shown by the local computers as the workload increased. High frame rates do not necessarily correspond to a better user experience, and for many participants, the more consistent the frame rate, the better the experience, depending on their expectations and needs.
    \item Table \ref{tab:pricecomparison} shows that other than the maximum use case, where an environment is used continuously for four years, the cloud environments provide competitive alternatives when compared to purchasing mid-to-high end laptops and workstations.  They also offer a degree of flexibility that increases their financial suitability.
    \item Being powerful enough for the task required does not only mean achieving the maximum, but also achieving a suitable minimum.  A local computer chosen for its suitability in a high compute demand scenario, will be greatly overpowered for standard office operations, such as checking email or editing \LaTeX{} documents.  This investigation has shown that cloud environments may be better suited to the tasks, as presented, than the local computers.
    \item High availability is critical to the business practices of cloud providers, and research network infrastructure is just as critical to research institutions. Networking performance during the tasks showed that a reasonably stable network with moderate bandwidth is sufficient to use a VHD.  This type of networking is available at most research institutions in Australia, where our user experience testing was conducted, and much of the world.
\end{enumerate}


\section*{Acknowledgements}

We thank the participants for contributing their time to this project. All experimental work was approved and conducted in accordance with the requirements of Swinburne University Human Research Ethics Committee (SUHREC).  We also thank Anita, Amy and Jacob Meade for their assistance in preparation and data entry of the participant results, Dr Dany Vohl (Swinburne University of Technology) for his support with Shwirl, and Dr Glenn Kacprzak (Swinburne University of Technology) for sample image data.  We also thank Dr Stephen Giugni and Dr Steven Manos (University of Melbourne) for supporting this research, and Terry Brennan (University of Melbourne) for assisting with the costing of the Nectar virtual machines, Dr David Perry and Dylan McCullough (University of Melbourne) for providing the build script for the cloud virtual machines. This research was supported by use of the Nectar Research Cloud and by the Melbourne Node at the University of Melbourne.  The Nectar Research Cloud is a collaborative Australian research platform supported by the National Collaborative Research Infrastructure Strategy.  This research was also supported by Adrian While and Craig Lawton through the "AWS Cloud Credits for Research" program from Amazon Web Services.

Funding for the Sloan Digital Sky Survey IV has been provided by the Alfred P. Sloan Foundation, the U.S. Department of Energy Office of Science, and the Participating Institutions. SDSS-IV acknowledges
support and resources from the Center for High-Performance Computing at
the University of Utah. The SDSS web site is www.sdss.org.

SDSS-IV is managed by the Astrophysical Research Consortium for the 
Participating Institutions of the SDSS Collaboration including the 
Brazilian Participation Group, the Carnegie Institution for Science, 
Carnegie Mellon University, the Chilean Participation Group, the French Participation Group, Harvard-Smithsonian Center for Astrophysics, 
Instituto de Astrof\'isica de Canarias, The Johns Hopkins University, 
Kavli Institute for the Physics and Mathematics of the Universe (IPMU) / 
University of Tokyo, Lawrence Berkeley National Laboratory, 
Leibniz Institut f\"ur Astrophysik Potsdam (AIP),  
Max-Planck-Institut f\"ur Astronomie (MPIA Heidelberg), 
Max-Planck-Institut f\"ur Astrophysik (MPA Garching), 
Max-Planck-Institut f\"ur Extraterrestrische Physik (MPE), 
National Astronomical Observatories of China, New Mexico State University, 
New York University, University of Notre Dame, 
Observat\'ario Nacional / MCTI, The Ohio State University, 
Pennsylvania State University, Shanghai Astronomical Observatory, 
United Kingdom Participation Group,
Universidad Nacional Aut\'onoma de M\'exico, University of Arizona, 
University of Colorado Boulder, University of Oxford, University of Portsmouth, 
University of Utah, University of Virginia, University of Washington, University of Wisconsin, 
Vanderbilt University, and Yale University.




\bibliographystyle{model2-names}\biboptions{authoryear}
\bibliography{VHD} 

\begin{thebibliography}{39}
\expandafter\ifx\csname natexlab\endcsname\relax\def\natexlab#1{#1}\fi
\expandafter\ifx\csname url\endcsname\relax
  \def\url#1{\texttt{#1}}\fi
\expandafter\ifx\csname urlprefix\endcsname\relax\def\urlprefix{URL }\fi
\providecommand{\eprint}[2][]{\url{#2}}
\providecommand{\bibinfo}[2]{#2}
\ifx\xfnm\relax \def\xfnm[#1]{\unskip,\space#1}\fi
\bibitem[{Abolfathi et~al.(2017)Abolfathi, Aguado, Aguilar, Prieto, Almeida,
  Ananna, Anders, Anderson, Andrews, Anguiano, Aragon-Salamanca,
  Argudo-Fernandez, Armengaud, Ata, Aubourg, Avila-Reese, Badenes, Bailey,
  Balland, Barger, Barrera-Ballesteros, Bartosz, Bastien, Bates, Baumgarten,
  Bautista, Beaton, Beers, Belfiore, Bender, Bernardi, Bershady, Beutler, Bird,
  Bizyaev, Blanc, Blanton, Blomqvist, Bolton, Boquien, Borissova, Bovy, Diaz,
  Brandt, Brinkmann, Brownstein, Bundy, Burgasser, Burtin, Busca, Canas,
  Cano-Diaz, Cappellari, Carrera, Casey, Sodi, Chen, Cherinka, Chiappini, Choi,
  Chojnowski, Chuang, Chung, Clerc, Cohen, Comerford, Comparat, Nascimento,
  da~Costa, Cousinou, Covey, Crane, Cruz-Gonzalez, Cunha, Ilha, Damke, Darling,
  Davidson~Jr., Dawson, Lizaola, de~la Macorra, de~la Torre, De~Lee, Agathe,
  Machado, Dell'Agli, Delubac, Diamond-Stanic, Donor, Downes, Drory, Bourboux,
  Duckworth, Dwelly, Dyer, Ebelke, Eigenbrot, Eisenstein, Elsworth, Emsellem,
  Eracleous, Erfanianfar, Escoffier, Fan, Alvar, Fernandez-Trincado, Cirolini,
  Feuillet, Finoguenov, Fleming, Font-Ribera, Freischlad, Frinchaboy, Fu, Chew,
  Galbany, Perez, Garcia-Dias, Garcia-Hernandez, Oehmichen, Gaulme, Gelfand,
  Gil-Marin, Gillespie, Goddard, Hernandez, Gonzalez-Perez, Grabowski, Green,
  Grier, Gueguen, Guo, Guy, Hagen, Hall, Harding, Hasselquist, Hawley, Hayes,
  Hearty, Hekker, Hernandez, Toledo, Hogg, Holley-Bockelmann, Holtzman, Hou,
  Hsieh, Hunt, Hutchinson, Hwang, Angel, Johnson, Jones, Jonsson, Jullo, Khan,
  Kinemuchi, Kirkby, Kirkpatrick~IV, Kitaura, Knapp, Kneib, Kollmeier, Lacerna,
  Lane, Lang, Law, Goff, Lee, Li, Li, Lian, Liang, Lima, Lin, Long, Lucatello,
  Lundgren, Mackereth, MacLeod, Mahadevan, Maia, Majewski, Manchado, Maraston,
  Mariappan, Marques-Chaves, Masseron, Masters, McDermid, McGreer, Melendez,
  Meneses-Goytia, Merloni, Merrifield, Meszaros, Meza, Minchev, Minniti,
  Mueller, Muller-Sanchez, Muna, Munoz, Myers, Nair, Nandra, Ness, Newman,
  Nichol, Nidever, Nitschelm, Noterdaeme, O'Connell, Oelkers, Oravetz, Oravetz,
  Ortiz, Osorio, Pace, Padilla, Palanque-Delabrouille, Palicio, Pan, Pan,
  Parikh, Paris, Park, Peirani, Pellejero-Ibanez, Penny, Percival,
  Perez-Fournon, Petitjean, Pieri, Pinsonneault, Pisani, Prada, Prakash,
  Queiroz, Raddick, Raichoor, Rembold, Richstein, Riffel, Riffel, Rix, Robin,
  Torres, Roman-Zuniga, Ross, Rossi, Ruan, Ruggeri, Ruiz, Salvato, Sanchez,
  Sanchez, Almeida, Sanchez-Gallego, Rojas, Santiago, Schiavon, Schimoia,
  Schlafly, Schlegel, Schneider, Schuster, Schwope, Seo, Serenelli, Shen, Shen,
  Shetrone, Shull, Aguirre, Simon, Skrutskie, Slosar, Smethurst, Smith, Sobeck,
  Somers, Souter, Souto, Spindler, Stark, Stassun, Steinmetz, Stello,
  Storchi-Bergmann, Streblyanska, Stringfellow, Suarez, Sun, Szigeti,
  Taghizadeh-Popp, Talbot, Tang, Tao, Tayar, Tembe, Teske, Thaker, Thomas,
  Tissera, Tojeiro, Tremonti, Troup, Urry, Valenzuela, Bosch, Vargas-Gonzalez,
  Vargas-Magana, Vazquez, Villanova, Vogt, Wake, Wang, Weaver, Weijmans,
  Weinberg, Westfall, Whelan, Wilcots, Wild, Williams, Wilson, Wood-Vasey,
  Wylezalek, Xiao, Yan, Yang, Ybarra, Yeche, Zakamska, Zamora, Zarrouk,
  Zasowski, Zhang, Zhao, Zhao, Zheng, Zheng, Zhou, Zhu, Zinn \&
  Zou}]{abolfathi_fourteenth_2017}
\bibinfo{author}{Abolfathi, B.}, \bibinfo{author}{Aguado, D.S.},
  \bibinfo{author}{Aguilar, G.}, \bibinfo{author}{Prieto, C.A.},
  \bibinfo{author}{Almeida, A.}, et~al., \bibinfo{year}{2017}.
\newblock \bibinfo{title}{The {Fourteenth} {Data} {Release} of the {Sloan}
  {Digital} {Sky} {Survey}: {First} {Spectroscopic} {Data} from the extended
  {Baryon} {Oscillation} {Spectroscopic} {Survey} and from the second phase of
  the {Apache} {Point} {Observatory} {Galactic} {Evolution} {Experiment}}.
\newblock \bibinfo{journal}{arXiv:1707.09322 [astro-ph]} \bibinfo{note}{ArXiv:
  1707.09322}.
\bibitem[{Astrocompute(2015)}]{astrocompute_seeing_2015}
\bibinfo{author}{Astrocompute}, \bibinfo{year}{2015}.
\newblock \bibinfo{title}{Seeing stars through the {Cloud} - {SKA}
  {Telescope}}.
\newblock \bibinfo{note}{Available at:
  https://skatelescope.org/news/ska-aws-astrocompute-cloud-computing-grant/}.
\bibitem[{Ball(2013)}]{ball_canfar+_2013}
\bibinfo{author}{Ball, N.M.}, \bibinfo{year}{2013}.
\newblock \bibinfo{title}{{CANFAR}+ {Skytree}: {A} {Cloud} {Computing} and
  {Data} {Mining} {System} for {Astronomy}}.
\newblock \bibinfo{journal}{arXiv preprint arXiv:1312.3996} .
\bibitem[{Berriman \& Groom(2011)}]{berriman_how_2011}
\bibinfo{author}{Berriman, B.}, \bibinfo{author}{Groom, S.L.},
  \bibinfo{year}{2011}.
\newblock \bibinfo{title}{How will astronomy archives survive the data
  tsunami?}
\newblock \bibinfo{journal}{Communications of the ACM} \bibinfo{volume}{54},
  \bibinfo{pages}{52}.
\bibitem[{Berriman et~al.(2012)Berriman, Brinkworth, Gelino, Wittman, Deelman,
  Juve, Rynge \& Kinney}]{berriman_tale_2012}
\bibinfo{author}{Berriman, G.B.}, \bibinfo{author}{Brinkworth, C.},
  \bibinfo{author}{Gelino, D.}, \bibinfo{author}{Wittman, D.K.},
  \bibinfo{author}{Deelman, E.}, et~al., \bibinfo{year}{2012}.
\newblock \bibinfo{title}{A {Tale} {Of} 160 {Scientists}, {Three}
  {Applications}, {A} {Workshop} and {A} {Cloud}}.
\newblock \bibinfo{journal}{arXiv preprint arXiv:1211.4055} .
\bibitem[{Berriman et~al.(2013)Berriman, Deelman, Juve, Rynge \&
  V\"ockler}]{berriman_application_2013}
\bibinfo{author}{Berriman, G.B.}, \bibinfo{author}{Deelman, E.},
  \bibinfo{author}{Juve, G.}, \bibinfo{author}{Rynge, M.},
  \bibinfo{author}{V\"ockler, J.S.}, \bibinfo{year}{2013}.
\newblock \bibinfo{title}{The application of cloud computing to scientific
  workflows: a study of cost and performance}.
\newblock \bibinfo{journal}{Philosophical Transactions of the Royal Society A:
  Mathematical, Physical and Engineering Sciences} \bibinfo{volume}{371}.
\bibitem[{Berriman \& Good(2017)}]{berriman_application_2017}
\bibinfo{author}{Berriman, G.B.}, \bibinfo{author}{Good, J.C.},
  \bibinfo{year}{2017}.
\newblock \bibinfo{title}{The {Application} of the {Montage} {Image} {Mosaic}
  {Engine} to the {Visualization} of {Astronomical} {Images}}.
\newblock \bibinfo{journal}{Publications of the Astronomical Society of the
  Pacific} \bibinfo{volume}{129}, \bibinfo{pages}{058006}.
\bibitem[{Berriman et~al.(2016)Berriman, Good, Rusholme \&
  Robitaille}]{berriman_next_2016}
\bibinfo{author}{Berriman, G.B.}, \bibinfo{author}{Good, J.C.},
  \bibinfo{author}{Rusholme, B.}, \bibinfo{author}{Robitaille, T.},
  \bibinfo{year}{2016}.
\newblock \bibinfo{title}{The {Next} {Generation} of the {Montage} {Image}
  {Mosaic} {Toolkit}}.
\newblock \bibinfo{journal}{arXiv preprint arXiv:1608.02649} .
\bibitem[{Berriman et~al.(2010)Berriman, Juve, Deelman, Regelson \&
  Plavchan}]{berriman_application_2010}
\bibinfo{author}{Berriman, G.B.}, \bibinfo{author}{Juve, G.},
  \bibinfo{author}{Deelman, E.}, \bibinfo{author}{Regelson, M.},
  \bibinfo{author}{Plavchan, P.}, \bibinfo{year}{2010}.
\newblock \bibinfo{title}{The application of cloud computing to astronomy: {A}
  study of cost and performance}, in: \bibinfo{booktitle}{e-{Science}
  {Workshops}, 2010 {Sixth} {IEEE} {International} {Conference} on}, pp.
  \bibinfo{pages}{1--7}.
\bibitem[{Bevan(2009)}]{bevan_usability_2009}
\bibinfo{author}{Bevan, N.}, \bibinfo{year}{2009}.
\newblock \bibinfo{title}{Usability}, in: \bibinfo{editor}{Liu, L.},
  \bibinfo{editor}{\"Ozsu, M.T.} (Eds.), \bibinfo{booktitle}{Encyclopedia of
  {Database} {Systems}}. \bibinfo{publisher}{Springer US},
  \bibinfo{address}{Boston, MA}, pp. \bibinfo{pages}{3247--3251}.
\bibitem[{Bipinchandra et~al.(2014)Bipinchandra, Aluvalu \&
  Shanker~Singh}]{bipinchandra_intelligent_2014}
\bibinfo{author}{Bipinchandra, G.K.}, \bibinfo{author}{Aluvalu, R.},
  \bibinfo{author}{Shanker~Singh, A.}, \bibinfo{year}{2014}.
\newblock \bibinfo{title}{Intelligent {Resource} {Allocation} {Technique} for
  {Desktop}-as-a-{Service} in {Cloud} {Environment}}.
\newblock \bibinfo{journal}{International Journal of Computer Applications}
  \bibinfo{volume}{96}, \bibinfo{pages}{43--48}.
\bibitem[{Borne(2008)}]{kargupta_scientific_2008}
\bibinfo{author}{Borne, K.}, \bibinfo{year}{2008}.
\newblock \bibinfo{title}{Scientific {Data} {Mining} in {Astronomy}}, in:
  \bibinfo{editor}{Kargupta, H.}, \bibinfo{editor}{Han, J.},
  \bibinfo{editor}{Yu, P.}, \bibinfo{editor}{Motwani, R.},
  \bibinfo{editor}{Kumar, V.} (Eds.), \bibinfo{booktitle}{Next {Generation} of
  {Data} {Mining}}. \bibinfo{publisher}{Chapman and Hall/CRC}. volume
  \bibinfo{volume}{20084157}.
\bibitem[{Brunner et~al.(2002)Brunner, Djorgovski, Prince \&
  Szalay}]{abello_massive_2002}
\bibinfo{author}{Brunner, R.J.}, \bibinfo{author}{Djorgovski, S.G.},
  \bibinfo{author}{Prince, T.A.}, \bibinfo{author}{Szalay, A.S.},
  \bibinfo{year}{2002}.
\newblock \bibinfo{title}{Massive {Datasets} in {Astronomy}}, in:
  \bibinfo{editor}{Abello, J.}, \bibinfo{editor}{Pardalos, P.M.},
  \bibinfo{editor}{Resende, M.G.C.} (Eds.), \bibinfo{booktitle}{Handbook of
  {Massive} {Data} {Sets}}. \bibinfo{publisher}{Springer US},
  \bibinfo{address}{Boston, MA}. volume~\bibinfo{volume}{4}, pp.
  \bibinfo{pages}{931--979}.
\bibitem[{Caton \& Hawkins(2009)}]{caton_remote_2009}
\bibinfo{author}{Caton, D.B.}, \bibinfo{author}{Hawkins, L.},
  \bibinfo{year}{2009}.
\newblock \bibinfo{title}{Remote {Observing}: {Equipment}, {Methods} and
  {Experiences} at the {Dark} {Sky} {Observatory}}, in:
  \bibinfo{booktitle}{American {Astronomical} {Society} {Meeting} {Abstracts}
  \#213}, p. \bibinfo{pages}{427}.
\bibitem[{Deboosere et~al.(2012)Deboosere, Vankeirsbilck, Simoens, De~Turck,
  Dhoedt \& Demeester}]{deboosere_cloud-based_2012}
\bibinfo{author}{Deboosere, L.}, \bibinfo{author}{Vankeirsbilck, B.},
  \bibinfo{author}{Simoens, P.}, \bibinfo{author}{De~Turck, F.},
  \bibinfo{author}{Dhoedt, B.}, et~al., \bibinfo{year}{2012}.
\newblock \bibinfo{title}{Cloud-{Based} {Desktop} {Services} for {Thin}
  {Clients}}.
\newblock \bibinfo{journal}{IEEE Internet Computing} \bibinfo{volume}{16},
  \bibinfo{pages}{60--67}.
\bibitem[{Deelman et~al.(2008)Deelman, Singh, Livny, Berriman \&
  Good}]{deelman_cost_2008}
\bibinfo{author}{Deelman, E.}, \bibinfo{author}{Singh, G.},
  \bibinfo{author}{Livny, M.}, \bibinfo{author}{Berriman, B.},
  \bibinfo{author}{Good, J.}, \bibinfo{year}{2008}.
\newblock \bibinfo{title}{The cost of doing science on the cloud: the montage
  example}, in: \bibinfo{booktitle}{Proceedings of the 2008 {ACM}/{IEEE}
  conference on {Supercomputing}}, \bibinfo{publisher}{IEEE Press}.
  p.~\bibinfo{pages}{50}.
\bibitem[{Dodson et~al.(2016)Dodson, Vinsen, Wu, Popping, Meyer, Wicenec,
  Quinn, van Gorkom \& Momjian}]{dodson_imaging_2016}
\bibinfo{author}{Dodson, R.}, \bibinfo{author}{Vinsen, K.},
  \bibinfo{author}{Wu, C.}, \bibinfo{author}{Popping, A.},
  \bibinfo{author}{Meyer, M.}, et~al., \bibinfo{year}{2016}.
\newblock \bibinfo{title}{Imaging {SKA}-scale data in three different computing
  environments}.
\newblock \bibinfo{journal}{Astronomy and Computing} \bibinfo{volume}{14},
  \bibinfo{pages}{8--22}.
\bibitem[{Duato et~al.(1997)Duato, Yalamanchili \&
  Ni}]{duato_interconnection_1997}
\bibinfo{author}{Duato, J.}, \bibinfo{author}{Yalamanchili, S.},
  \bibinfo{author}{Ni, L.M.}, \bibinfo{year}{1997}.
\newblock \bibinfo{title}{Interconnection networks: an engineering approach}.
\newblock \bibinfo{publisher}{IEEE Computer Society Press},
  \bibinfo{address}{Los Alamitos, Calif}.
\bibitem[{Goscinski et~al.(2015)Goscinski, Paterson, Hines, McIntosh, Thompson,
  Ryan, Hall, Maksimenko, Bambery \& Panjikar}]{goscinski_massive:_2015}
\bibinfo{author}{Goscinski, W.J.}, \bibinfo{author}{Paterson, D.},
  \bibinfo{author}{Hines, C.}, \bibinfo{author}{McIntosh, P.},
  \bibinfo{author}{Thompson, D.}, et~al., \bibinfo{year}{2015}.
\newblock \bibinfo{title}{{MASSIVE}: an {HPC} {Collaboration} to {Underpin}
  {Synchrotron} {Science}}, in: \bibinfo{booktitle}{Proceedings of
  {ICALEPCS}2015}, \bibinfo{address}{Melbourne, Australia}.
\bibitem[{Hassan \& Fluke(2011)}]{hassan_scientific_2011}
\bibinfo{author}{Hassan, A.}, \bibinfo{author}{Fluke, C.},
  \bibinfo{year}{2011}.
\newblock \bibinfo{title}{Scientific {Visualization} in {Astronomy}: {Towards}
  the {Petascale} {Astronomy} {Era}}.
\newblock \bibinfo{journal}{Publications of the Astronomical Society of
  Australia} \bibinfo{volume}{28}, \bibinfo{pages}{150--170}.
\bibitem[{Hoffa et~al.(2008)Hoffa, Mehta, Freeman, Deelman, Keahey, Berriman \&
  Good}]{hoffa_use_2008}
\bibinfo{author}{Hoffa, C.}, \bibinfo{author}{Mehta, G.},
  \bibinfo{author}{Freeman, T.}, \bibinfo{author}{Deelman, E.},
  \bibinfo{author}{Keahey, K.}, et~al., \bibinfo{year}{2008}.
\newblock \bibinfo{title}{On the use of cloud computing for scientific
  workflows}, in: \bibinfo{booktitle}{{eScience}, 2008. {eScience}'08. {IEEE}
  {Fourth} {International} {Conference} on}, \bibinfo{publisher}{IEEE}. pp.
  \bibinfo{pages}{640--645}.
\bibitem[{Hong et~al.(2017)Hong, Spence \& Nikolopoulos}]{hong_gpu_2017}
\bibinfo{author}{Hong, C.H.}, \bibinfo{author}{Spence, I.},
  \bibinfo{author}{Nikolopoulos, D.S.}, \bibinfo{year}{2017}.
\newblock \bibinfo{title}{{GPU} {Virtualization} and {Scheduling} {Methods}:
  {A} {Comprehensive} {Survey}}.
\newblock \bibinfo{journal}{ACM Computing Surveys} \bibinfo{volume}{50},
  \bibinfo{pages}{1--37}.
\bibitem[{Iserte et~al.(2016)Iserte, Clemente-Castello, Castello, Mayo \&
  Quintana-Orti}]{cardoso_closer_2016}
\bibinfo{author}{Iserte, S.}, \bibinfo{author}{Clemente-Castello, F.},
  \bibinfo{author}{Castello, A.}, \bibinfo{author}{Mayo, R.},
  \bibinfo{author}{Quintana-Orti, E.}, \bibinfo{year}{2016}.
\newblock \bibinfo{title}{{CLOSER} 2016: proceedings of the 6th {International}
  {Conference} on {Cloud} {Computing} and {Services} {Science}: {Rome},
  {Italy}, {April} 23-25, 2016}.
\newblock \bibinfo{publisher}{SCITEPRESS - Science and Technology Publications,
  Lda}, \bibinfo{address}{Setúbal, Portugal}.
\bibitem[{Juric \& Tyson(2012)}]{juric_lsst_2012}
\bibinfo{author}{Juric, M.}, \bibinfo{author}{Tyson, T.}, \bibinfo{year}{2012}.
\newblock \bibinfo{title}{{LSST} {Data} {Management}: {Entering} the {Era} of
  {Petascale} {Optical} {Astronomy}}.
\newblock \bibinfo{journal}{Proceedings of the International Astronomical
  Union} \bibinfo{volume}{10}, \bibinfo{pages}{675--676}.
\bibitem[{Juve et~al.(2009)Juve, Deelman, Vahi, Mehta, Berriman, Berman \&
  Maechling}]{juve_scientific_2009}
\bibinfo{author}{Juve, G.}, \bibinfo{author}{Deelman, E.},
  \bibinfo{author}{Vahi, K.}, \bibinfo{author}{Mehta, G.},
  \bibinfo{author}{Berriman, B.}, et~al., \bibinfo{year}{2009}.
\newblock \bibinfo{title}{Scientific workflow applications on {Amazon} {EC}2},
  in: \bibinfo{booktitle}{E-{Science} {Workshops}, 2009 5th {IEEE}
  {International} {Conference} on}, \bibinfo{publisher}{IEEE}. pp.
  \bibinfo{pages}{59--66}.
\bibitem[{Khalid et~al.(2016)Khalid, Shoaib, Sarfraz, Shabbir, Shaheed \&
  Rubab}]{khalid_desktop_2016}
\bibinfo{author}{Khalid, F.}, \bibinfo{author}{Shoaib, U.},
  \bibinfo{author}{Sarfraz, M.S.}, \bibinfo{author}{Shabbir, A.},
  \bibinfo{author}{Shaheed, S.M.}, et~al., \bibinfo{year}{2016}.
\newblock \bibinfo{title}{Desktop {Virtualization}: {An} {Art} to {Manage} and
  {Maintain} affordable {PC} infrastructure}.
\newblock \bibinfo{journal}{International Journal of Computer Science and
  Information Security} \bibinfo{volume}{14}, \bibinfo{pages}{187}.
\bibitem[{Lam et~al.(2012)Lam, Bertini, Isenberg, Plaisant \&
  Carpendale}]{lam_empirical_2012}
\bibinfo{author}{Lam, H.}, \bibinfo{author}{Bertini, E.},
  \bibinfo{author}{Isenberg, P.}, \bibinfo{author}{Plaisant, C.},
  \bibinfo{author}{Carpendale, S.}, \bibinfo{year}{2012}.
\newblock \bibinfo{title}{Empirical {Studies} in {Information} {Visualization}:
  {Seven} {Scenarios}}.
\newblock \bibinfo{journal}{IEEE Transactions on Visualization and Computer
  Graphics} \bibinfo{volume}{18}, \bibinfo{pages}{1520--1536}.
\bibitem[{Massimino et~al.(2014)Massimino, Costa, Becciani, Vuerli,
  Bandieramonte, Petta, Riggi, Sciacca, Vitello \&
  Pistagna}]{massimino_acid_2014}
\bibinfo{author}{Massimino, P.}, \bibinfo{author}{Costa, A.},
  \bibinfo{author}{Becciani, U.}, \bibinfo{author}{Vuerli, C.},
  \bibinfo{author}{Bandieramonte, M.}, et~al., \bibinfo{year}{2014}.
\newblock \bibinfo{title}{{ACID} {Astronomical} and {Physics} {Cloud}
  {Interactive} {Desktop}: {A} {Prototype} of {VUI} for {CTA} {Science}
  {Gateway}}, in: \bibinfo{editor}{Manset, N.}, \bibinfo{editor}{Forshay, P.}
  (Eds.), \bibinfo{booktitle}{Astronomical {Data} {Analysis} {Software} and
  {Systems} {XXIII}}, p. \bibinfo{pages}{293}.
\bibitem[{Meade et~al.(2013)Meade, Manos, Sinnott, Fluke, Knijff \&
  Tseng}]{meade_research_2013}
\bibinfo{author}{Meade, B.}, \bibinfo{author}{Manos, S.},
  \bibinfo{author}{Sinnott, R.}, \bibinfo{author}{Fluke, C.},
  \bibinfo{author}{Knijff, D.v.d.}, et~al., \bibinfo{year}{2013}.
\newblock \bibinfo{title}{Research {Cloud} {Data} {Communities}}, in:
  \bibinfo{booktitle}{{THETA} 2013}, \bibinfo{address}{Hobart, Tasmania}.
\newblock \bibinfo{note}{Copyright 2013 THETA: The Higher Education Technology
  Agenda}.
\bibitem[{Miller \& Pegah(2007)}]{miller_virtualization:_2007}
\bibinfo{author}{Miller, K.}, \bibinfo{author}{Pegah, M.},
  \bibinfo{year}{2007}.
\newblock \bibinfo{title}{Virtualization: virtually at the desktop},
  \bibinfo{publisher}{ACM Press}. pp. \bibinfo{pages}{255--260}.
\bibitem[{Nieh et~al.(2000)Nieh, Yang \& Novik}]{nieh_comparison_2000}
\bibinfo{author}{Nieh, J.}, \bibinfo{author}{Yang, S.J.},
  \bibinfo{author}{Novik, N.}, \bibinfo{year}{2000}.
\newblock \bibinfo{title}{A comparison of thin-client computing architectures}.
\newblock \bibinfo{type}{Technical Report}. Technical Report CUCS-022-00,
  Department of Computer Science, Columbia University.
\bibitem[{Rampersad et~al.(2017)Rampersad, Blyth, Elson \&
  Kuttel}]{rampersad_improving_2017}
\bibinfo{author}{Rampersad, L.}, \bibinfo{author}{Blyth, S.},
  \bibinfo{author}{Elson, E.}, \bibinfo{author}{Kuttel, M.M.},
  \bibinfo{year}{2017}.
\newblock \bibinfo{title}{Improving the usability of scientific software with
  participatory design: a new interface design for radio astronomy
  visualisation software}, \bibinfo{publisher}{ACM Press}. pp.
  \bibinfo{pages}{1--9}.
\bibitem[{Ravi et~al.(2011)Ravi, Becchi, Agrawal \&
  Chakradhar}]{ravi_supporting_2011}
\bibinfo{author}{Ravi, V.T.}, \bibinfo{author}{Becchi, M.},
  \bibinfo{author}{Agrawal, G.}, \bibinfo{author}{Chakradhar, S.},
  \bibinfo{year}{2011}.
\newblock \bibinfo{title}{Supporting {GPU} sharing in cloud environments with a
  transparent runtime consolidation framework}, in:
  \bibinfo{booktitle}{Proceedings of the 20th international symposium on {High}
  performance distributed computing}, \bibinfo{publisher}{ACM}. pp.
  \bibinfo{pages}{217--228}.
\bibitem[{Rimal et~al.(2010)Rimal, Choi \& Lumb}]{antonopoulos_taxonomy_2010}
\bibinfo{author}{Rimal, B.P.}, \bibinfo{author}{Choi, E.},
  \bibinfo{author}{Lumb, I.}, \bibinfo{year}{2010}.
\newblock \bibinfo{title}{A {Taxonomy}, {Survey}, and {Issues} of {Cloud}
  {Computing} {Ecosystems}}, in: \bibinfo{editor}{Antonopoulos, N.},
  \bibinfo{editor}{Gillam, L.} (Eds.), \bibinfo{booktitle}{Cloud {Computing}}.
  \bibinfo{publisher}{Springer London}, \bibinfo{address}{London}, pp.
  \bibinfo{pages}{21--46}.
\bibitem[{Sabater et~al.(2017)Sabater, S\'anchez-Exp\'osito, Best, Garrido,
  Verdes-Montenegro \& Lezzi}]{sabater_calibration_2017}
\bibinfo{author}{Sabater, J.}, \bibinfo{author}{S\'anchez-Exp\'osito, S.},
  \bibinfo{author}{Best, P.}, \bibinfo{author}{Garrido, J.},
  \bibinfo{author}{Verdes-Montenegro, L.}, et~al., \bibinfo{year}{2017}.
\newblock \bibinfo{title}{Calibration of {LOFAR} data on the cloud}.
\newblock \bibinfo{journal}{Astronomy and computing} \bibinfo{volume}{19},
  \bibinfo{pages}{75--89}.
\bibitem[{V\"ockler et~al.(2011)V\"ockler, Juve, Deelman, Rynge \&
  Berriman}]{vockler_experiences_2011}
\bibinfo{author}{V\"ockler, J.S.}, \bibinfo{author}{Juve, G.},
  \bibinfo{author}{Deelman, E.}, \bibinfo{author}{Rynge, M.},
  \bibinfo{author}{Berriman, B.}, \bibinfo{year}{2011}.
\newblock \bibinfo{title}{Experiences using cloud computing for a scientific
  workflow application}, in: \bibinfo{booktitle}{Proceedings of the 2nd
  international workshop on {Scientific} cloud computing},
  \bibinfo{publisher}{ACM}. pp. \bibinfo{pages}{15--24}.
\bibitem[{Vohl(2017)}]{vohl_shwirl:_2017}
\bibinfo{author}{Vohl, D.}, \bibinfo{year}{2017}.
\newblock \bibinfo{title}{Shwirl: {Meaningful} coloring of spectral cube data
  with volume rendering}.
\newblock \bibinfo{note}{Published: Astrophysics Source Code Library}.
\bibitem[{Walter et~al.(2008)Walter, Brinks, de~Blok, Bigiel, Kennicutt,
  Thornley \& Leroy}]{walter_things:_2008}
\bibinfo{author}{Walter, F.}, \bibinfo{author}{Brinks, E.},
  \bibinfo{author}{de~Blok, W.J.G.}, \bibinfo{author}{Bigiel, F.},
  \bibinfo{author}{Kennicutt, R.C.}, et~al., \bibinfo{year}{2008}.
\newblock \bibinfo{title}{{THINGS}: {THE} {H} {I} {NEARBY} {GALAXY} {SURVEY}}.
\newblock \bibinfo{journal}{The Astronomical Journal} \bibinfo{volume}{136},
  \bibinfo{pages}{2563--2647}.
\bibitem[{Younge(2016)}]{younge_architectural_2016}
\bibinfo{author}{Younge, A.J.}, \bibinfo{year}{2016}.
\newblock \bibinfo{title}{Architectural principles and experimentation of
  distributed high performance virtual clusters}.
\newblock Ph.D. thesis. Indiana University.

\end{thebibliography}




\appendix
\section*{Appendix}
\renewcommand{\thesubsection}{\Alph{subsection}}
\renewcommand{\thefigure}{A\arabic{figure}}

\setcounter{figure}{0}


\subsection{Network}
\label{lbl:apndx-network}

The network performance for connections to the VHDs, while varied, delivers sufficient bandwidth and is robust enough to provide a user experience that matches the local desktop.

\begin{figure}
	\includegraphics[width=\columnwidth]{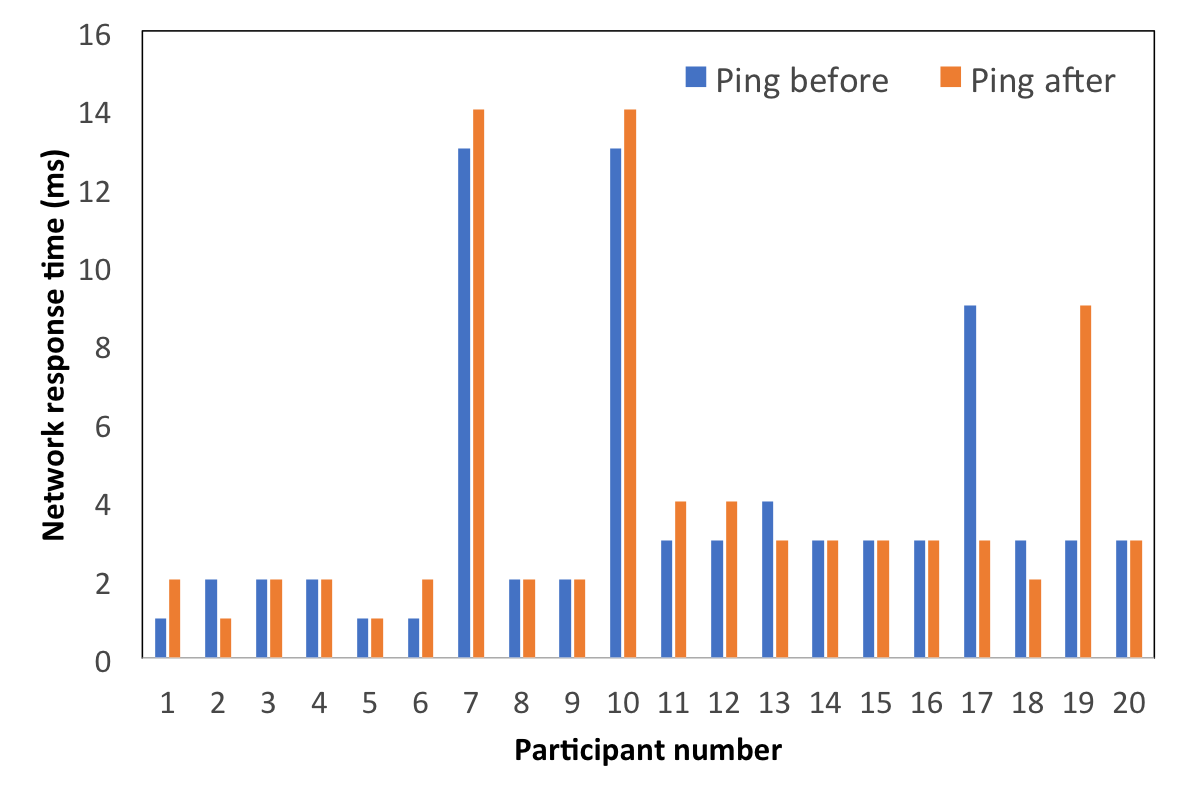}
    \caption{The network response time was measured before and after the tasks. The figure shows that the start and end states of the network environment did not vary greatly.  Despite the variation of ping response times, participants did not report any latency impact during their use of the VHDs.}
    \label{fig:ping}
\end{figure}

\begin{figure}
	\includegraphics[width=\columnwidth]{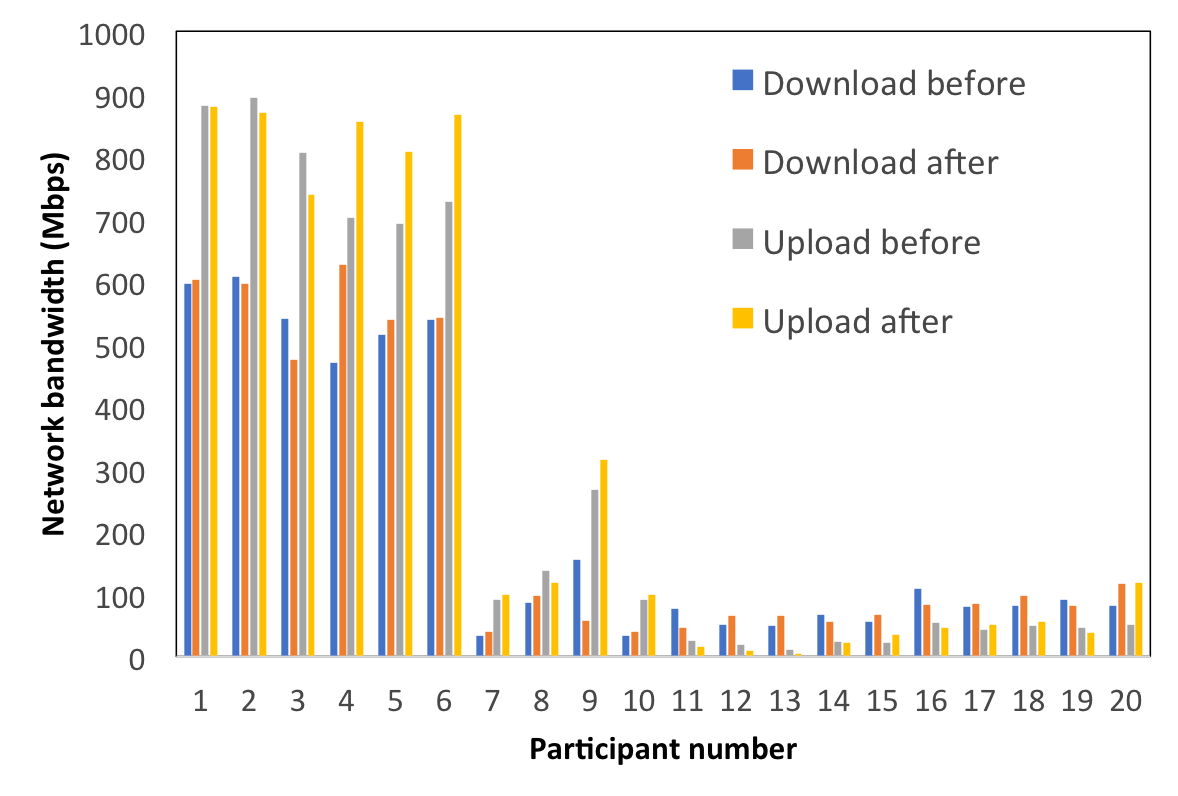}
    \caption{The network bandwidth for download and upload was measured before and after the tasks. Although they varied significantly due to the use of different connection types, no apparent impact on the performance of the cloud environments was identified by the participants.}
    \label{fig:network}
\end{figure}

Network Speedtests shown in Figures \ref{fig:ping} and \ref{fig:network} were conducted between July and September, 2017.

Despite the longer ping times for participants 7 and 10, and the larger variation in times for participants 17 and 19, the network did not affect the performance of the VHDs during those tasks.

\begin{figure}
	\includegraphics[width=\columnwidth]{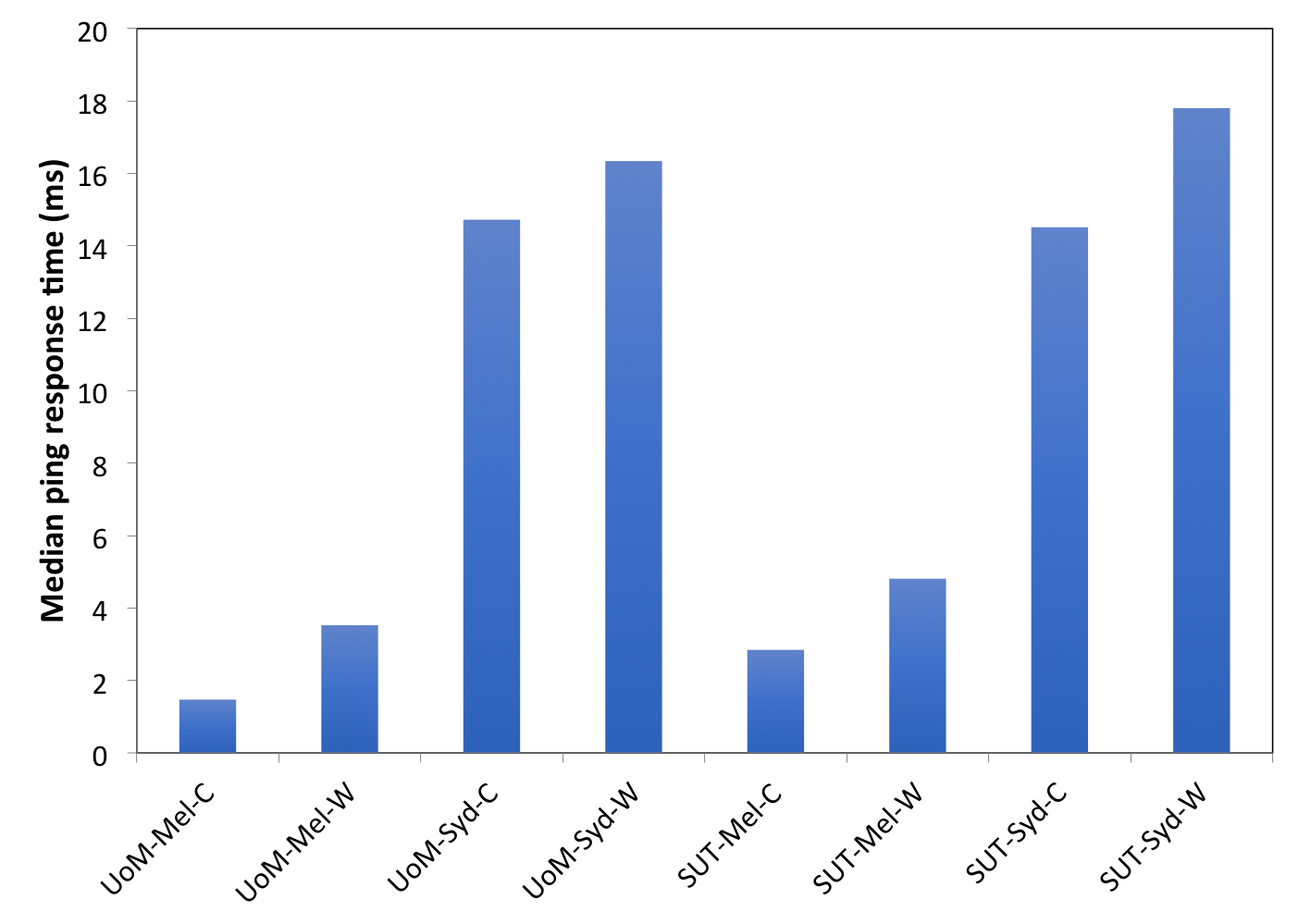}
    \caption{Network latency ping test results from study locations at Swinburne University of Technology (SUT) and the University of Melbourne (UoM) to the UoM data centre in Melbourne (Mel), Australia, and AWS data centre located in Sydney (Syd), Australia. C = laptop connected to network using a cable. W = laptop connected over wireless network.  Ping tests were conducted in March, 2018. }
    \label{fig:latencytest}
\end{figure}

Figure \ref{fig:latencytest} shows the network latency from the study locations to the virtual machines at the University of Melbourne data centre and the AWS Sydney data centre.  Included are the ping response times over a cable connected network and the wireless networks available at each testing site.

\subsection{User experience testing: procedure}
\label{lbl:apndx-exp-procedure}

The tasks completed during the user experience tests were intended to engage the participants in a manner that closely resembled typical astronomy use.  The most common experience is using windowed applications with a local mouse and keyboard, with information displayed on a local computer screen.  The design of the 2D phase of the investigation met that aim, by challenging the participant to use common astronomy software to complete a common astronomy task.  Using a clearly defined activity allowed the participants to focus their reflections on the usability of the environment in question.  This was preferable to having the participants freely explore the interface, which might have yielded more `operating system'-centric evaluations.

The purpose of the 3D phase of the investigation was to test the capacity of the environments under graphically intensive load.  Many astronomy tasks involve the use of 3D models and volume, so it is imperative to understand how VHDs handle these workloads, and if they are comparable to local GPU computation.

Each user testing session took 35-45 minutes depending on the participant.  This consisted of:
\begin{enumerate}
    \item Introduction and pre-investigation interview (5 minutes)
    \item 2D image alignment activity (10 minutes)
    \item 3D spectral cube manipulation activity (15 minutes)
    \item Post-investigation interview (5 minutes)
\end{enumerate}

The 2D image alignment procedure was as follows:
\begin{enumerate}
    \item Start DS9
    \item Load first image
    \item Load second image in new frame buffer
    \item Observe the offset using the Blink mode
    \item Start IRAF
    \item Using the \texttt{imshift} command, create a shifted image
    \item Load the shifted image into the second frame buffer in DS9
    \item Observe the lack of offset using the Blink mode
\end{enumerate}

The 3D spectral cube manipulation procedure was as follows:
\begin{enumerate}
    \item Start Shwirl and load the spectral cube
    \item Switch to the Camera and Adjustment tab and set the Z-Scale to `10'.  Rotate the volume using the mouse, and then reset the Z-Scale to `1'.
    \item Switch to the Colour tab and choose a new colour scheme for the volume.  Rotate the volume using the mouse and then reset the colour scheme (Note: some participants did not reset the colour scheme, but no performance difference was observed in the results).
    \item Switch to the Filter tab and adjust the High and Low filters.  Rotate the volume and then reset the values.
    \item Switch to the Smoothing tab, and set the Smoothing value to 3.  Rotate the volume.
    \item Repeat the previous step with smoothing values of 5, 7, 9, 11 and 13.
\end{enumerate}

\label{lastpage}
\end{document}